\documentclass[11pt, journal, onecolumn]{IEEEtran}

\usepackage{url}
\usepackage{hyperref}
\usepackage{amsthm}
\usepackage{multirow}
\usepackage{todonotes}

\usepackage[capitalize,noabbrev]{cleveref}
\usepackage{nth} % Legacy Packages
\usepackage{booktabs} % Legacy Packages

\hypersetup{
  colorlinks = true,
  citecolor  = blue,
  urlcolor   = blue,
  linkcolor  = blue
}

\theoremstyle{plain}
\newtheorem{thm}{Theorem}
\theoremstyle{definition}
\newtheorem{defn}[thm]{Definition}
\definecolor{gray}{rgb}{0.5, 0.5, 0.5}
\newcommand{\yy}{\textcolor{blue}{\textbf{yes}}}
\newcommand{\nn}{\textcolor{gray}{\textbf{no}}}

\newcommand{\tabMonitorControlAction}[1]{
\small
\begin{tabular}{@{}cccc@{}}
\toprule
\textbf{Magnitude} & \textbf{Cause}                                                                          & \textbf{Adverse Effect}                                                    & \textbf{Control Action}                                             \\ \midrule
Temperature        & \begin{tabular}[c]{@{}c@{}}Localized\\ power\\ consumption\end{tabular}                 & \begin{tabular}[c]{@{}c@{}}Performance /\\ Reliability\\ loss\end{tabular} & \begin{tabular}[c]{@{}c@{}}DVFS /\\ Clock\\ throttling\end{tabular} \\
Critical Path      & \begin{tabular}[c]{@{}c@{}}Timing\\ uncertainties\end{tabular}                          & Speed loss                                                                 & DVFS                                                                \\
Soft Error         & \begin{tabular}[c]{@{}c@{}}ncreasing in\\ system\\ operating\\ frequencies\end{tabular} & \begin{tabular}[c]{@{}c@{}}Reliability\\ loss\end{tabular}                 & \begin{tabular}[c]{@{}c@{}}Backward\\ recovery\end{tabular}         \\
Aging              & \begin{tabular}[c]{@{}c@{}}Prolonged\\ Usage\end{tabular}                               & \begin{tabular}[c]{@{}c@{}}Speed\\ reduction/\\ Malfunction\end{tabular}   & DVFS                                                                \\ \bottomrule
\end{tabular}
}

\newcommand{\tabOldSurveyA}{
\begin{table}[!ht]
\caption{Comparison among the reviewed surveys related to aging monitor types, and this survey coverage. Legend: ``A'': addressed, ``NA'':  not addressed.}
\label{tab:comparison_monitor}
\centering
\small
\begin{tabular}{@{}lccc@{}}
\toprule
\multicolumn{1}{l}{\textbf{}}                                    & \multicolumn{1}{l}{\textbf{~\cite{rahimipour2012survey}}} & \multicolumn{1}{l}{\textbf{~\cite{kochte2017self}}} &
\textbf{This Survey} \\ \midrule
\textbf{Temperature}                                             & A                                    & A                                    & A                                                              \\
\textbf{\begin{tabular}[c]{@{}r@{}}Critical Path: Timing Errors\end{tabular}} & NA                                   & A                                    & A                                                              \\
\textbf{\begin{tabular}[c]{@{}r@{}}Critical Path: SEU\end{tabular}}   & A                                    & A                                    & A                                                              \\
\textbf{Critical Path: Delay}                                                   & A                                    & NA                                   & A                                                              \\
\textbf{Clock Frequency}                                               & NA                                   & NA                                   & A                                                              \\
\textbf{Workload}                                                & NA                                   & A                                    & A                                                              \\
\textbf{\begin{tabular}[c]{@{}r@{}}Circuit State\end{tabular}} & NA                                   & NA                                   & A                                                              \\
\textbf{Voltage}                                                 & NA                                   & NA                                   & A                                                              \\ \bottomrule
\end{tabular}
\end{table}
}

\newcommand{\tabOldSurveyB}[1]{
\begin{table}[#1]
\caption{Comparison among the reviewed surveys related to reconfiguration techniques types, and this survey coverage. Legend: ``A'': addressed, ``NA'':  not addressed.}
\label{tab:comparison_reconf}
\centering
\small
\begin{tabular}{@{}lccc@{}}
\toprule
\multicolumn{1}{l}{\textbf{}}                                              & \multicolumn{1}{l}{\textbf{~\cite{rahimipour2012survey}}} & \multicolumn{1}{l}{\textbf{~\cite{khoshavi2017contemporary}}} & \textbf{This Survey} \\ \midrule
\textbf{\begin{tabular}[c]{@{}r@{}}Dynamic Voltage Scaling\end{tabular}}        & A                                    & A                                    & A                                                              \\
\textbf{\begin{tabular}[c]{@{}r@{}}Dynamic Frequency Scaling\end{tabular}}       & A                                    & A                                    & A                                                              \\
\textbf{\begin{tabular}[c]{@{}r@{}}Aging Compensation\end{tabular}}      & NA                                   & A                                    & A                                                              \\
\textbf{\begin{tabular}[c]{@{}r@{}}Body Bias Adaptive\end{tabular}}      & NA                                   & NA                                   & A                                                              \\
\textbf{\begin{tabular}[c]{@{}r@{}}Workload Reduction\end{tabular}}      & NA                                   & NA                                   & A                                                              \\
\textbf{\begin{tabular}[c]{@{}r@{}}Clock Throttling\end{tabular}}        & A                                    & NA                                   & NA                                                             \\
\textbf{\begin{tabular}[c]{@{}r@{}}Backward Recovering\end{tabular}}     & A                                    & NA                                   & NA                                                             \\
\textbf{\begin{tabular}[c]{@{}r@{}}Computational Sprinting\end{tabular}} & NA                                   & A                                    & NA                                                             \\ \bottomrule
\end{tabular}
\end{table}
}

\newcommand{\tabCriticalPath}[1]{
\begin{table}[#1]
\caption{Comparison among critical-path monitors papers. Legend: ``A'': addressed, ``NA'':  not addressed.}
\label{tab:path_delay_summary}
\centering
\small
\begin{tabular}{@{}rccccc@{}}
\toprule
\multicolumn{1}{l}{}         & \textbf{System} & \textbf{Extra} & \textbf{Delay} & \textbf{Comparison} \\ 
\multicolumn{1}{l}{}         & \textbf{monitoring} & \textbf{clock} & \textbf{element} & \textbf{mechanism} \\ \midrule
\textbf{\cite{savanur2015bist}}, 2015        & NA          & NA          & A       & A        \\
\textbf{\cite{sai2017cost}}, 2017            & NA          & NA          & A       & A        \\
\textbf{\cite{sai2018multi}}, 2018          & NA          & NA          & A       & A        \\
\textbf{\cite{copetti2016nbti}}, 2016        & NA          & NA          & A       & A        \\
\textbf{\cite{sadeghi2015online}}, 2015      & NA          & A           & NA      & A        \\
\textbf{\cite{kamalself}}, 2017              & NA          & A           & NA      & A        \\
\textbf{\cite{yi2014scan}}, 2014             & NA          & NA          & NA      & A        \\
\textbf{\cite{jung2017diagnosing}}, 2017     & NA          & A           & NA      & A        \\
\textbf{\cite{vazquez2014error}}, 2014       & NA          & NA          & NA      & A        \\
\textbf{\cite{chandra2014monitoring}}, 2014  & NA          & NA          & A       & A        \\
\textbf{\cite{masuda2018mttf}}, 2018         & NA          & NA          & A       & A        \\
\textbf{\cite{vijayan2017workload}}, 2017    & NA          & NA          & NA      & A        \\
\textbf{\cite{saliva2015digital}}, 2015      & NA          & NA          & A       & A        \\
\textbf{\cite{di2019hidden}}, 2019           & NA          & NA          & NA      & A        \\
\textbf{\cite{wang2019aging}}, 2019          & NA          & NA          & NA      & A        \\
\textbf{\cite{rohbani2018low}}, 2018         & NA          & NA          & A       & NA       \\
\textbf{\cite{pachito2012aging}}, 2012       & A           & NA          & A       & A        \\
\textbf{\cite{cho2015aging}}, 2015           & NA          & NA          & NA      & NA       \\
\textbf{\cite{ding2016design}}, 2016         & NA          & NA          & NA      & A        \\
\textbf{\cite{li2014robust}}, 2014           & NA          & NA          & NA      & NA       \\
\textbf{\cite{jang2014chip}}, 2014           & NA          & NA          & A       & A        \\ \bottomrule
\end{tabular}
\end{table}
}

\newcommand{\tabSumary}[1]{
\begin{table}[#1]
\caption{Summary of the literature review.}
\label{tab:literature_summary}
\centering
\small
\begin{tabular}{@{}ccccc@{}}
\toprule
                  & \textbf{Overall} & \textbf{Monitor Insertion} & \textbf{Structure} & \textbf{Metastability} \\ 
                  & \textbf{Monitoring} & \textbf{Strategy} & \textbf{Reuse} & \textbf{Concern} \\ \midrule
\textbf{\cite{igarashi2015digital}}, 2015         & \yy          & \nn            & \nn           & \nn     \\
\textbf{\cite{majerus2017embedded}}, 2017         & \yy          & \nn            & \nn           & \nn     \\
\textbf{\cite{sengupta2017estimating}}, 2017      & \yy          & \nn            & \nn           & \nn     \\
\textbf{\cite{kim2017investigation}}, 2017        & \yy          & \nn            & \nn           & \nn     \\
\textbf{\cite{shakya2015performance}}, 2015       & \yy          & \nn            & \nn           & \nn     \\
\textbf{\cite{miyake2016temperature}}, 2016       & \yy          & \nn            & \nn           & \nn     \\
\textbf{\cite{kumar2014chip}}, 2014               & \yy          & \nn            & \nn           & \nn     \\
\textbf{\cite{ali2016accessing}}, 2016            & \yy          & \nn            & \nn           & \nn     \\
\textbf{\cite{rathore2019lifeguard}}, 2019        & \yy          & \nn            & \nn           & \nn     \\
\textbf{\cite{savanur2015bist}}, 2015             & \nn          & \nn            & \yy           & \nn     \\
\textbf{\cite{sai2017cost}}, 2017                 & \nn          & \nn            & \nn           & \nn     \\
\textbf{\cite{sai2018multi}}, 2018                & \nn          & \nn            & \nn           & \yy     \\
\textbf{\cite{copetti2016nbti}}, 2016             & \nn          & \nn            & \nn           & \nn     \\
\textbf{\cite{sadeghi2015online}}, 2015           & \yy          & \nn            & \nn           & \nn     \\
\textbf{\cite{kamalself}}, 2017                   & \nn          & \nn            & \nn           & \nn     \\
\textbf{\cite{yi2014scan}}, 2014                  & \nn          & \nn            & \nn           & \nn     \\
\textbf{\cite{jung2017diagnosing}}, 2017          & \nn          & \nn            & \nn           & \nn     \\
\textbf{\cite{vazquez2014error}}, 2014            & \nn          & \yy            & \nn           & \nn     \\
\textbf{\cite{chandra2014monitoring}}, 2014       & \nn          & \yy            & \nn           & \nn     \\
\textbf{\cite{masuda2018mttf}}, 2018              & \nn          & \nn            & \nn           & \nn     \\
\textbf{\cite{vijayan2017workload}}, 2017         & \nn          & \yy            & \nn           & \nn     \\
\textbf{\cite{saliva2015digital}}, 2015           & \nn          & \yy            & \nn           & \nn     \\
\textbf{\cite{di2019hidden}}, 2019                & \yy          & \nn            & \nn           & \nn     \\
\textbf{\cite{wang2019aging}}, 2019               & \yy          & \yy            & \nn           & \nn     \\
\textbf{\cite{rohbani2018low}}, 2018              & \nn          & \nn            & \nn           & \nn     \\
\textbf{\cite{pachito2012aging}}, 2012            & \nn          & \yy            & \nn           & \nn     \\
\textbf{\cite{semiao2014aging}}, 2014             & \nn          & \yy            & \nn           & \nn     \\
\textbf{\cite{semiao2014performance}}, 2014       & \nn          & \yy            & \nn           & \nn     \\
\textbf{\cite{cho2015aging}}, 2015                & \yy          & \nn            & \nn           & \nn     \\
\textbf{\cite{ding2016design}}, 2016              & \yy          & \nn            & \nn           & \nn     \\
\textbf{\cite{li2014robust}}, 2014                & \nn          & \nn            & \nn           & \nn     \\
\textbf{\cite{jang2014chip}}, 2014                & \nn          & \yy            & \nn           & \nn     \\
\textbf{\cite{wang2015aging}}, 2015               & \yy          & \nn            & \nn           & \yy     \\
\textbf{\cite{kufluoglu2017thermally}}, 2017      & \yy          & \nn            & \nn           & \nn     \\
\textbf{\cite{koneru2015fine}}, 2015              & \yy          & \nn            & \yy           & \nn     \\
\textbf{\cite{firouzi2015re}}, 2015               & \yy          & \nn            & \yy           & \nn     \\
\textbf{\cite{khan2009self}}, 2019                & \yy          & \nn            & \nn           & \nn     \\
\textbf{\cite{sadi2017design}}, 2017              & \yy          & \nn            & \yy           & \nn     \\
\textbf{\cite{baranowski2015line}}, 2015          & \yy          & \nn            & \nn           & \nn     \\
\textbf{\cite{narang2015nbti}}, 2015              & \yy          & \nn            & \nn           & \nn     \\ 
\textbf{\cite{li2018linear}}, 2018                & \yy          & \nn            & \nn           & \nn     \\ \bottomrule
\end{tabular}
\end{table}
}

\newcommand{\tabMonitors}[1]{
\begin{table}[#1]
\centering
\small
\caption{Classification of the monitors' works according to the proposed classification.}
\label{tab:monitor_class}
\begin{tabular}{lcccccc}
\hline
\textbf{Types}               & \multicolumn{6}{c}{Works}     \\ \hline

Temperature                  & \cite{igarashi2015digital} & \cite{majerus2017embedded} & \cite{sengupta2017estimating} & \cite{kim2017investigation} & \cite{shakya2015performance} & \cite{miyake2016temperature} \\
                             & \cite{kumar2014chip} & \cite{ali2016accessing} & \cite{rathore2019lifeguard} & \cite{semiao2014aging} & \cite{semiao2014performance} \\

Critical path: timing errors & \cite{savanur2015bist} & \cite{sai2017cost} & \cite{sai2018multi} & \cite{copetti2016nbti} & \cite{sadeghi2015online} & \cite{kamalself} \\ 
                             & \cite{yi2014scan} & \cite{jung2017diagnosing} & \cite{vazquez2014error} & \cite{chandra2014monitoring} & \cite{masuda2018mttf} & \cite{vijayan2017workload} \\
                             & \cite{saliva2015digital}  &  \cite{di2019hidden} &  \cite{wang2019aging}  \\

Critical path: single event upset       & \cite{rohbani2018low}  \\
Critical path: delay         & \cite{pachito2012aging} & \cite{cho2015aging} & \cite{ding2016design} & \cite{li2014robust} & \cite{jang2014chip}   \\
Clock frequency              & \cite{wang2015aging} & \cite{kufluoglu2017thermally}   \\
Circuit state                & \cite{koneru2015fine} & \cite{firouzi2015re} & \cite{khan2009self} & \cite{sadi2017design}  \\
Workload                     & \cite{baranowski2015line}  \\
Voltage                      & \cite{narang2015nbti} & \cite{li2018linear}   \\ \hline
\end{tabular}
\end{table}
}

\newcommand{\tabReconfig}[1]{
\begin{table}[#1]
\centering
\small
\caption{Works classification regarding reconfiguration techniques.}
\label{tab:reconf_class}
\begin{tabular}{lccccccc}
\toprule
\textbf{Techiniques} & \multicolumn{7}{c}{Works} \\ \midrule
Voltage scaling      & \cite{igarashi2015digital} & \cite{copetti2016nbti} & \cite{masuda2018mttf} & \cite{pachito2012aging} & \cite{cho2015aging} & \cite{li2014robust}  & \cite{khan2009self} \\
Frequency scaling    & \cite{igarashi2015digital} & \cite{pachito2012aging} & \cite{wang2015aging} & \cite{khan2009self}   \\
Aging compensation   & \cite{kumar2014chip}   \\
Body bias adjust     & \cite{narang2015nbti}    \\
Workload reduction   & \cite{vijayan2017workload}   \\ \bottomrule
\end{tabular}
\end{table}
}

\newcommand{\tabConclusion}[1]{
\begin{table}[#1]
\centering
\small
\caption{Classification according to the implementation method.}
\label{tab:class}
\begin{tabular}{ll}
\toprule
\textbf{Classification}                    & \textbf{Works} \\ 
\midrule
Digital, Hardware, Synchronous & \cite{sengupta2017estimating}~\cite{kim2017investigation}~\cite{shakya2015performance}~\cite{miyake2016temperature}~\cite{kumar2014chip}~\cite{rathore2019lifeguard}~\cite{savanur2015bist} \\
							   & \cite{sai2017cost}~\cite{sai2018multi}~\cite{copetti2016nbti}~\cite{sadeghi2015online}~\cite{kamalself}~\cite{yi2014scan}~\cite{jung2017diagnosing} \\
							   & \cite{vazquez2014error}~\cite{chandra2014monitoring}~\cite{masuda2018mttf}~\cite{di2019hidden}~\cite{wang2019aging}~\cite{rohbani2018low}~\cite{pachito2012aging} \\
                               & \cite{semiao2014aging}~\cite{semiao2014performance}~\cite{cho2015aging}~\cite{ding2016design}~\cite{li2014robust}~\cite{jang2014chip}~\cite{wang2015aging} \\ 
                               & \cite{kufluoglu2017thermally}~\cite{firouzi2015re}~\cite{li2018linear} \\
Digital, Hardware \& Software, Synchronous & \cite{igarashi2015digital}~\cite{vijayan2017workload}~\cite{koneru2015fine}~\cite{khan2009self}~\cite{sadi2017design}~\cite{baranowski2015line} \\
Analog \& Digital, Hardware, Synchronous   & \cite{majerus2017embedded}~\cite{ali2016accessing}~\cite{saliva2015digital}~\cite{narang2015nbti} \\ \bottomrule
\end{tabular}
\end{table}
}

\begin{document}

\title{A Survey of Aging Monitors and Reconfiguration Techniques}

\author{\IEEEauthorblockN{Leonardo R. Juracy, Matheus Trevisan Moreira, Alexandre de Morais Amory, Fernando Gehm Moraes}
\IEEEauthorblockA{PUCRS – Av. Ipiranga 6681, Porto Alegre, Brazil --  fernando.moraes@pucrs.br}}

\maketitle

\begin{abstract}
CMOS technology scaling makes aging effects an important concern for the design and fabrication of integrated circuits. Aging deterioration reduces the useful life of a circuit, making it fail earlier. This deterioration can affect all portions of a circuit and impacts its performance and reliability. Contemporary literature shows solutions to monitor and mitigate aging using hardware and software monitoring mechanisms and reconfiguration techniques. The goal of this review of the state-of-the-art is to identify existing monitoring and reconfiguration solutions for aging. This survey evaluates the aging research, focusing the years from 2012 to 2019, and proposes a classification for monitors and reconfiguration techniques. Results show that the most common monitor type used for aging detection is to monitor timing errors, and the most common reconfiguration technique used to deal with aging is voltage scaling. Furthermore, most of the literature contributions are in the digital field, using hardware solutions for monitoring aging in circuits. There are few literature contributions in the analog area, being the scope of this survey in the digital domain. By scrutinizing these solutions, this survey points directions for further research and development of aging monitors and reconfiguration techniques.
\end{abstract}


%%%%%%%%%%%%%%%%%%%%%%%%%%%%%%%%%%%%%%%%%%%%%%%%%%%%%%%%%%%%%%%
%% Introduction
%%%%%%%%%%%%%%%%%%%%%%%%%%%%%%%%%%%%%%%%%%%%%%%%%%%%%%%%%%%%%%%
\section{Introduction}
\label{sec:intro}

With the scaling of CMOS technology circuit reliability issues become increasingly relevant for the design of integrated circuits. Among these issues, circuit aging is getting critical, as it inevitably affects all circuits. The literature shows an increase in the number of papers related to aging between 2000 to 2020 \cite{kim2020aging, sai2020cost, sahoo2020novel}. This trend is due to the development of ultra-deep submicron technologies, where reliability became crucial \cite{lu2009statistical}.

Aging is the deterioration of circuit performance over time \cite{agarwal2008optimized}, which can reduce the useful life of a circuit. This deterioration can increase circuit delay and affect all portions of a System-on-Chip (SoC), analog circuit, digital logic, and memory. One important factor that accelerates aging is power. The increase of power in modern circuits increases the temperature and makes these circuits susceptible to effects like bias temperature instability (BTI) and hot carrier injection (HCI)~\cite{baker2019cmos}.

One solution to deal with aging effects is to add design safety margins to the circuit, such as clock margins. These margins ensure correct operation even in the presence of aging effects once they are designed according to worst-case conditions. However, these margins decrease performance, as they increase the clock period. The literature presents other solutions to mitigate aging, allowing to extend the lifetime of chips. These solutions monitor parameters that indicate aging effects. The monitored parameters include temperature, frequency variation, delay variation, among others. Decision methods, as threshold voltage analysis, evaluate whether the monitored parameters have an appropriate range of values. Activation mechanisms, such as voltage and frequency adaptations, bring the circuit back to the safe parameters if these do not meet predefined constraints.

The \textit{goal} of this review is to study existing solutions for monitoring aging effects in circuits and dealing with them. This study seeks to answer two questions:  \textit{(i)} what solutions the literature presents for aging monitoring? and \textit{(ii)} what solutions the literature presents for circuit reconfiguration? This work surveys the literature from 2012 to 2019, while works \cite{rahimipour2012survey, khoshavi2017contemporary, kochte2017self} map aging monitors since 1998. Thus, this survey fills the gap related to aging surveys, covering the most recent works. 

The literature review considered three main search terms:
\begin{itemize}
    \item \textit{Integrated Circuits}: this category limits the search to the area of integrated circuits. The search string presents the terms used to reference this area, like VLSI, system, and architectures;
    
    \item \textit{Monitoring and Self-reconfigurable Circuits}: this category includes the monitoring area and the problems induced by aging, as Negative-Bias Temperature Instability (NBTI);
    
    \item \textit{Aging}: this category limits the research by aging problems.
\end{itemize}

The Scopus abstract and citation database (\url{https://www.scopus.com}) was the start point of the research. After that, the review researched in IEEE (\url{https://ieeexplore.ieee.org}) and ACM (\url{https://www.acm.org}) databases. Finally, the research focused on Google Scholar (\url{https://scholar.google.com}) database. The review considered journals and conference papers published since 2014.

This survey is organized as follows.  \Cref{sec:definitions} presents basic definitions to the understating of this work.  \Cref{sec:related_work} presents a review of other surveys available in the literature. \Cref{sec:prop_class} presents the proposed classification for the reviewed papers. \Cref{sec:discussion} discusses comparatively the reviewed works.  \Cref{sec:lit_review} presents in detail the literature survey. \Cref{sec:conc} concludes this survey.

%%%%%%%%%%%%%%%%%%%%%%%%%%%%%%%%%%%%%%%%%%%%%%%%%%%%%%%%%%%%%%%
%%% Basic Definitions
%%%%%%%%%%%%%%%%%%%%%%%%%%%%%%%%%%%%%%%%%%%%%%%%%%%%%%%%%%%%%%%
\section{Basic Definitions}
\label{sec:definitions}

This section provides definitions required for the understanding of this survey.

\begin{defn}\label{def_cp}
\textit{Critical path}: is the purely combinational path with the most significant delay between any two registers or primary I/O ports in a design. 
\end{defn}

\begin{defn}\label{def_se}
\textit{Soft-errors}:  errors that can occur in a circuit when it is exposed to radiation, like cosmic rays and neutrons particles. Single-event upset (SEU) is an example of soft-error, when the exposition to radiation can generate a change in the memory elements values, generating erroneous outputs \cite{jagirdar2007efficient}.
\end{defn}
    
Important effects are considered in the aging research: NBTI (Definition \ref{def_nbti}), PBTI (Definition \ref{def_pbti}), HCI (Definition \ref{def_hci}).

\begin{defn}\label{def_nbti}
\textit{NBTI} -- Negative Bias Temperature Instability: increase in the threshold voltage and consequently decrease drain current and transconductance of a transistor \cite{schroder2003negative}. It occurs in PMOS transistors and is caused by circuit aging.
\end{defn}

\begin{defn}\label{def_pbti}
\textit{PBTI} -- Positive Bias Temperature Instability: similar to NBTI, but in NMOS transistors. Since the introduction of the high-K gate dielectrics and metal gates transistors, the effect of PBTI becomes comparable to the NBTI one.
\end{defn}

\begin{defn}\label{def_hci}
\textit{HCI} -- Hot-Carrier Injection: consists of a voltage drop that produces a large electric field in the region near to the drain of a transistor in saturation mode. It can result in the change of transistor characteristics such as the threshold voltage \cite{maricau2011transistor}. According to Novak et al., from Intel,~\cite{novak2015transistor}, tri-gate technologies as 14nm can help to mitigate HCI effects toghter with body bias adjust techniques.
\end{defn}

Classic test approaches are adopted for aging monitoring and reconfiguration techniques:  DfT (Definition \ref{def_dft}), BIST (Definition \ref{def_bist}), ATPG (Definition \ref{def_atpg}).

\begin{defn}\label{def_dft}
\textit{DfT} -- Design for Testability: a set of techniques used to improve circuit testability by increasing controllability and observability in the design \cite{Book3}.
\end{defn}

\begin{defn}\label{def_bist}
\textit{BIST} -- Built-in Self-Test: a mechanism that tests the circuit itself, verifying all or a portion of the internal functionality of the design \cite{Book3}. The hardware and/or the software is built into integrated circuits allowing them to test their operation. The main advantage of BIST is the ability to test internal circuits having no direct connections to external pins or external testers. In addition, BIST allows testing in the field.
\end{defn}

\begin{defn}\label{def_atpg}
\textit{ATPG} --Automatic Test Pattern Generation: a method that finds an input sequence (called test vectors) that enables automatic test equipment (ATE) to distinguish between the correct circuit behavior and the faulty circuit behavior caused by defects.
\end{defn}

%%%%%%%%%%%%%%%%%%%%%%%%%%%%%%%%%%%%%%%%%%%%%%%%%%%%%%%%%%%%%%%
%%% Previous  Aging Surveys
%%%%%%%%%%%%%%%%%%%%%%%%%%%%%%%%%%%%%%%%%%%%%%%%%%%%%%%%%%%%%%%
\section{Previous Aging Surveys}
\label{sec:related_work}

\begin{figure}[!ht]
    \centering
    \includegraphics[width=0.5\linewidth]{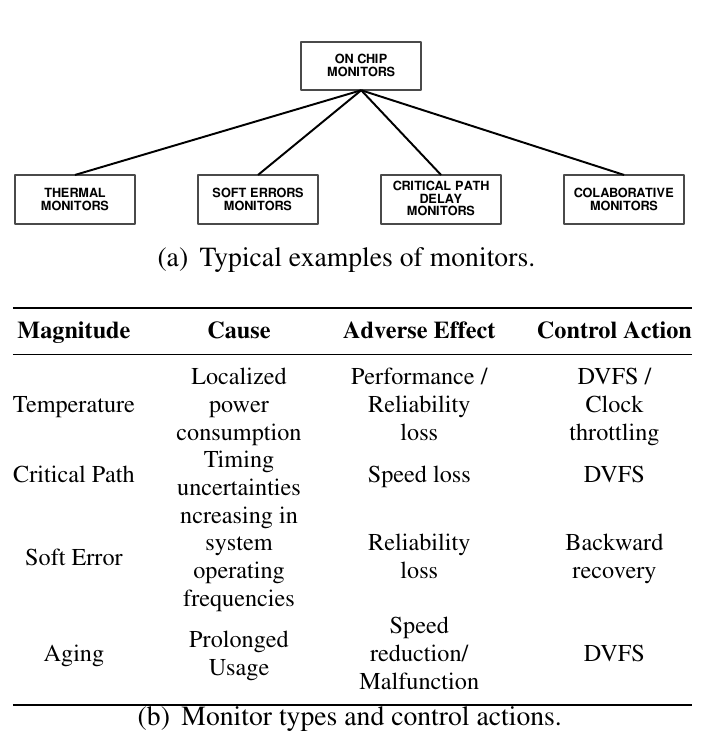}
    \caption{Monitors types and control actions. Adapted from \cite{rahimipour2012survey}.}
    \label{fig:2012_survey}
\end{figure}

Rahimipour et al. \cite{rahimipour2012survey} review on-chip monitors for temperature, soft-errors (Definition \ref{def_se}), and critical paths (Definition \ref{def_cp}), covering the period 1998-2011 (21 works). \Cref{fig:2012_survey}(a) presents their classification divided in: monitors for high temperature induced problems; monitors for soft-errors; monitors for critical paths; and collaborative monitors, which combines the previously mentioned class of monitors. Soft-errors detects change caused in memory elements due radiation. Thermal monitors identify problems induced by high temperatures. Critical path delay monitors detect changes in the critical path, as delay increases. Collaborative monitors are a combination of more than one class of monitors.  \Cref{fig:2012_survey}(b) summarizes the monitor types and also informs cause, effect, and control action for each effect.  

%
%   FIGURE 1 
%

\begin{figure}[!ht]
    \centering
    \includegraphics[width=1\linewidth]{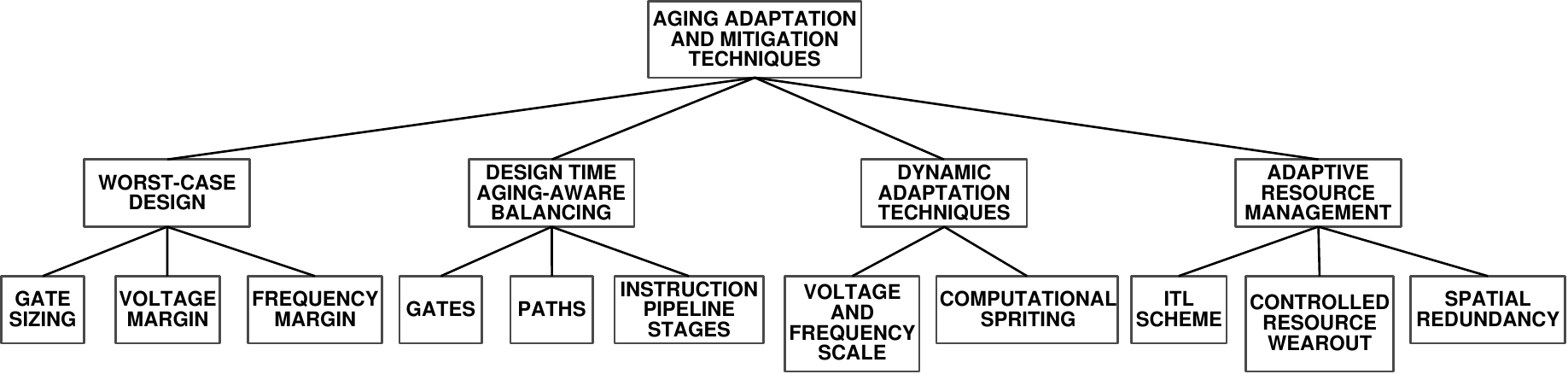}
    \caption{Taxonomy of aging mitigation techniques. Adapted from \cite{khoshavi2017contemporary}.}
    \label{fig:2017_taxonomy}
\end{figure}

Khoshavi et al. \cite{khoshavi2017contemporary} review the period 2004-2015 (30 works). Their work analyzes aging monitors and also aging models and techniques for aging mitigation. Aging models are used to predict the degradation of the circuit due to aging. Mitigation techniques are used to deal with aging effects and ensure correct behavior of the circuit under these effects. Besides, their work proposes a taxonomy to classify the aging mitigation techniques, shown in \Cref{fig:2017_taxonomy}. Worst-case design techniques add safety margins to the circuit characteristics, like frequency and supply voltage, at design-time. Design time aging-aware balancing focuses on balancing circuit delay to reduce aging effects. Dynamic adaptation techniques are online approaches to tune the design under aging during circuit operation. Adaptive resource management techniques mitigate the aging effects either through the management of idle time (Idle-Time Leveraging schemes, also called ITL schemes), power management and task scheduling (controlled resource wearout), or availability of expendable resources (spatial redundancy).

%
%   FIGURE 2
%

\begin{figure}[!ht]
    \centering
    \includegraphics[width=0.5\linewidth]{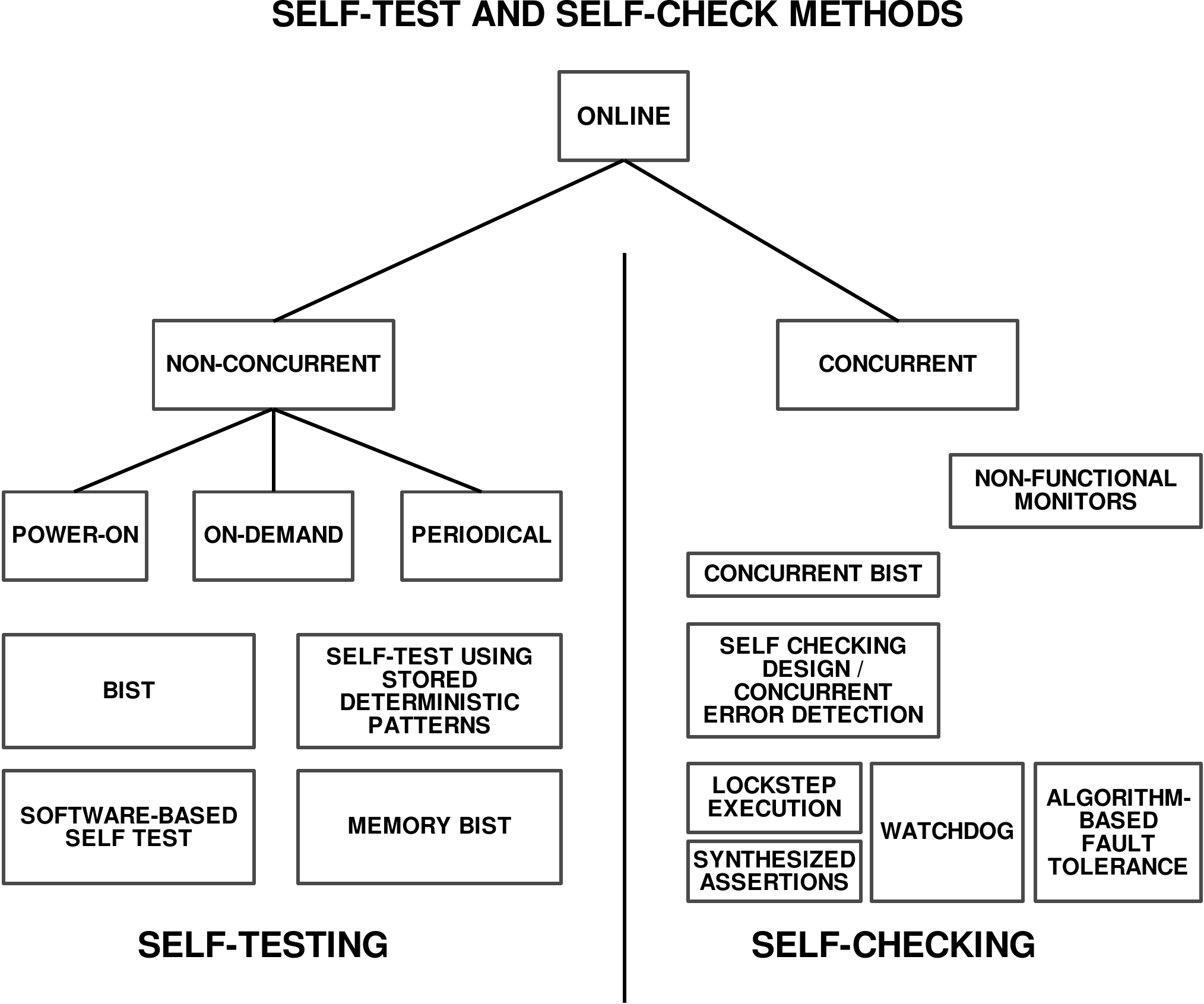}
    \caption{Self-test and self-checking circuits classification. Adapted from \cite{kochte2017self}.}
    \label{fig:2017_classification}
\end{figure}

Kochte et al. \cite{kochte2017self} review the period 2007-2017 (14 works). Their work analyses self-test, self-checking, and self-diagnosis online techniques for self-awareness. Online techniques include aging monitors. \Cref{fig:2017_classification} shows the classification for self-test and self-checking, which contains non-concurrent and concurrent approaches. Non-concurrent approaches regard self-testing methods, including classic methods, as BIST, and techniques like suspending the circuit operation to execute the test. Concurrent approaches are self-checking methods executed during the circuit operation, which includes aging monitors.

%
%   FIGURE 3 
%

%%%%%%%%%%%%%%%%%%%%%%%%%%%%%%%%%%%%%%%%%%%%%%%%%%%%%%%%%%%%%%%
%%% Proposed Classification
%%%%%%%%%%%%%%%%%%%%%%%%%%%%%%%%%%%%%%%%%%%%%%%%%%%%%%%%%%%%%%%
\section{Proposed Classification of Aging Monitors and Reconfiguration Techniques}
\label{sec:prop_class}

Our proposed classification extends the taxonomy of previous work \cite{rahimipour2012survey,khoshavi2017contemporary,kochte2017self}, considering parameters included in the state-of-the-art and features of the circuits monitored by them. Thus, this work proposes the following set of parameters for aging monitors classification:

\begin{itemize}
    \item \textit{Temperature}: a temperature monitor detects aging effects by measuring the variation in the temperature of the circuit. Temperature variations cause aging effects as NBTI and PBTI;
    
    \item \textit{Critical path}: critical path monitors can be used to identify timing errors due to wrong transitions, SEU, or delay increase on circuit paths. These errors are detectable by comparing a circuit path controlled by the system clock and the same path controlled by a delayed clock \cite{ernst2003razor};
    
    \item \textit{Clock frequency}: clock frequency monitors measure the frequency variation of a clock circuit under aging. This type of monitor uses a reference frequency to identify a change in the circuit operating frequency;
    
    \item \textit{Workload}: workload monitors measure the workload of a circuit to identify the stress level. Similar to the CPU load measure.
    
    \item \textit{Circuit state}: circuit state monitors consider the state of the entire system. For example, some techniques use the BIST circuit output to evaluate aging effects;
    
    \item \textit{Voltage}: Voltage monitors analyze the threshold voltage of the circuit. Aging effects as NBTI can change the threshold voltage of transistors.
\end{itemize}

\Cref{tab:comparison_monitor} presents the aging monitors addressed in this survey and compares with those covered by previous works. Note that the Table does not include survey \cite{khoshavi2017contemporary} because it does not consider aging monitors. Our work addresses eight monitors types, including monitors not covered by previous works.

%%%%%  tabela 1 - de comparação
\tabOldSurveyA

This work adopts the following classification of reconfiguration techniques:

\begin{itemize}
    \item \textit{Dynamic voltage scaling}: Dynamic Voltage Scaling (DVS) is a power management technique where the voltage used in a component increases or decreases, according to some criteria \cite{mittal2014survey}. An example of an approach to manage power dissipation is the adoption of a closed-loop control technique where the voltage is a knob to meet the power goal \cite{del2019hierarchical}. 
    
    \item \textit{Dynamic frequency scaling}: Dynamic frequency scaling (DFS) is a technique similar to DVS, but applied to the circuit frequency. Similarly, if the circuit needs a boost in performance, the frequency is increased. If the circuit or application can tolerate lower performance, the frequency may be decreased to allow power savings;
    
    \item \textit{Aging compensation}: Aging compensation is a technique that enables the circuit to alleviate aging effects. For example, some circuits activate extra devices, in parallel to the main circuit, to increase the driving strength of an output driver, compensating the degradation due to HCI and BTI effects~\cite{kumar2014chip};
    
    \item \textit{Body-bias adaptive}: Body-bias adaptive (BBA) is a technique that allows tuning the transistor threshold voltage \cite{chen2003comparison}. This technique helps to mitigate and compensate for the NBTI impact in the circuit;

    \item \textit{Workload reduction}: 
    Workload reduction is a series of software approaches to reduce the workload system by introducing, for instance, no operation (NOP) instructions during the system operation.

\end{itemize}

Table \ref{tab:comparison_reconf} compares reconfiguration approaches covered by the previous works, and the ones addressed in this survey. The Table does not include survey \cite{kochte2017self} because it does not address reconfiguration techniques. Also, note that this survey does not address three types of reconfiguration techniques (clock throttling, backward recovering and computational sprinting) that were covered in the previous surveys because these techniques were not adopted in the research papers from 2012 to 2019. Our work addresses five reconfiguration approaches, including two types not covered by previous works.

%%%%%  tabela 2 - de comparação
\tabOldSurveyB{!ht}

\section{Discussion related to the State-of-the-Art}
\label{sec:discussion}

This Section brings a summary of the presented techniques, remarks, and insights about how to deal with aging. Also, this Section answers the research questions presented in the Introduction Section. The classified state-of-the-art works are described in Section~\ref{sec:lit_review}.

About industry, most  applications rely on sensors built in SoCs that allow measuring variations in such parameters as the circuit ages. These sensors are commonly distributed across the die and accessible through DfT infrastructure or as peripherals to CPUs. The most common sensor consists of a set of ring oscillators that control asynchronous counters. These counters provide an overview of how the overall speed of the circuit is being impacted by aging \cite{wang2014silicon}.

%%%%%%%%%%%%%%%%%%%%%%%%%%%%%%%%%%%%%%%%%%%%%%%%%%%%%%%%%%%%%%%
%%%%%%%%%%%%%%%%%%%%%%%%%%%%%%%%%%%%%%%%%%%%%%%%%%%%%%%%%%%%%%%
\subsection{Summary and remarks of the literature review}

\Cref{tab:literature_summary} summarizes the reviewed works. The ``Overall Monitoring'' column means monitoring the entire circuit, not just the critical paths using, for example, the voltage or the current. The  ``Monitor Insertion Strategy'' column corresponds to approaches that use some strategy to insert the monitors, such as statistical methods, and not based only on critical paths. The ``Structure Reuse'' column shows designs that use structures available in the circuit for monitoring aging effects. In these cases, only DfT structures are reused. The ``Metastability Concern'' column contains works concerned with metastability issues.

%%%  tabela resumo
\tabSumary{!ht}

Circuits monitoring the overall system may present a low area overhead, once one mechanism can be applied for the entire design. This approach may be better in terms of area overhead when compared to solutions focusing on inserting monitors in all critical paths. However, designs that use methods to select paths to insert monitor are also promising in terms of area overhead reduction. The reuse of DfT structures is a promising strategy to choose paths, once test insertion overhead is already present in the circuit. This allows reducing the impact of aging monitors on area.

A "\textit{no}" in the first column (Overall Monitoring) means that only part of the system is monitored. This means that specialized mechanisms may be required for different parts of the system. Particularly, works with a "\textit{no}" in the second and/or third columns in \Cref{tab:literature_summary} (Monitor Insertion Strategy and Structure Reuse) can present a significant overhead in the system. Nevertheless, these solutions may be useful for detecting aging and could be combined with path selection strategies to reduce the system overhead, making their use feasible.  

Metastability is a concern present in only two works. It is an important issue, once its effects may propagate through designs, reducing circuit reliability, making them fail even in the absence of aging effects. Metastability can also be an issue to systems with more than one clock domain, due to the clock synchronization between domains. Thus, these  two works have an advantage  compared to other approaches since they use the same circuitry to deal with both aging and metastability effects.

A feature that can be observed is that voltage and frequency scaling are reconfiguration techniques associated mostly with critical path monitors, once it is possible to control the circuit speed by these two parameters. Similarly, body bias adjustment is associated with voltage monitors, once this reconfiguration technique changes the threshold voltage for compensating aging effects.

%%%%%%%%%%%%%%%%%%%%%%%%%%%%%%%%%%%%%%%%%%%%%%%%%%%%%%%%%%%%%%%
%%%%%%%%%%%%%%%%%%%%%%%%%%%%%%%%%%%%%%%%%%%%%%%%%%%%%%%%%%%%%%%
\subsection{Remarks about aging monitors}
\label{sec:monitor_result}

\begin{figure}[!ht]
    \centering
    \includegraphics[width=0.8\columnwidth]{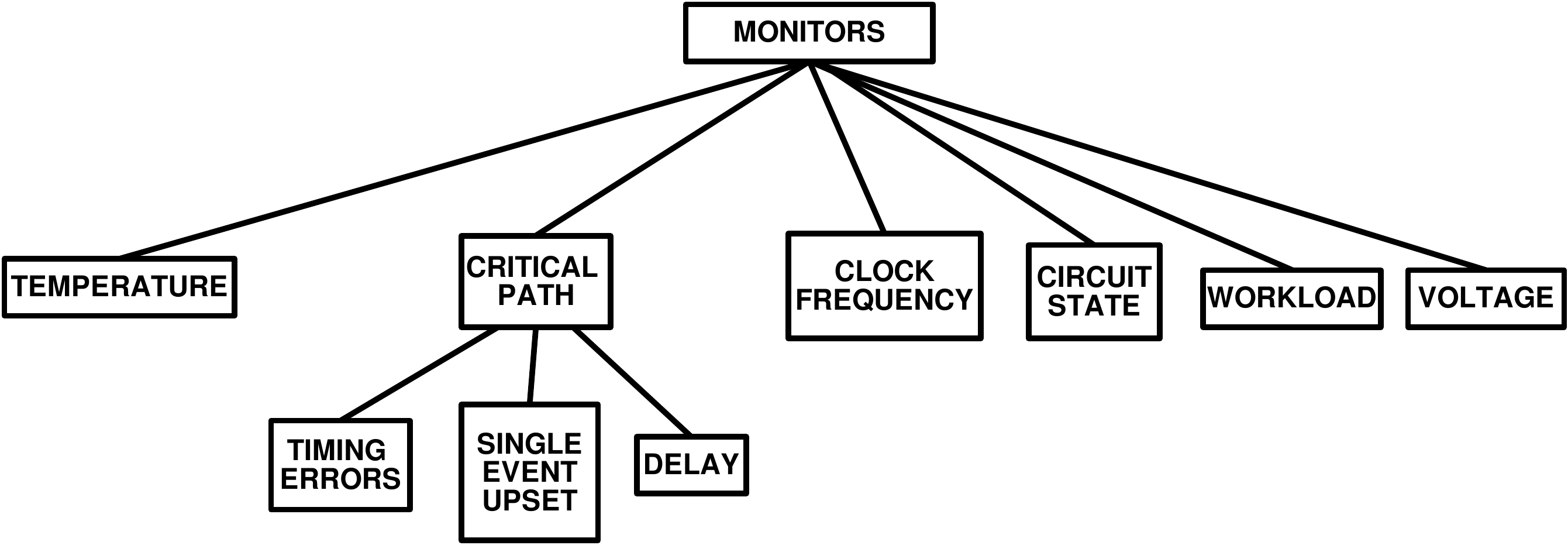}
    \caption{Monitor types covered by the survey.}
    \label{fig:monitor_tree}
\end{figure}

\Cref{fig:monitor_tree} and \Cref{tab:monitor_class} present the answer to the first research question of this survey, ``What solutions the literature presents for aging monitoring?''. The Figure shows the monitor types covered by this survey. The most common monitor type used for aging detection are timing error monitors, which is present in 36.58\% of the reviewed papers. The second most common monitor type used for aging detection is temperature monitor, which is present in 27\% of the papers.

%
%   FIGURE 4  
%
 
\tabMonitors{!ht}

%%%%%%%%%%%%%%%%%%%%%%%%%%%%%%%%%%%%%%%%%%%%%%%%%%%%%%%%%%%%%%%
%%%%%%%%%%%%%%%%%%%%%%%%%%%%%%%%%%%%%%%%%%%%%%%%%%%%%%%%%%%%%%%
\subsection{Remarks about reconfiguration techniques}
\label{sec:reconf_result}

\begin{figure}[!ht]
    \centering
    \includegraphics[width=0.7\columnwidth]{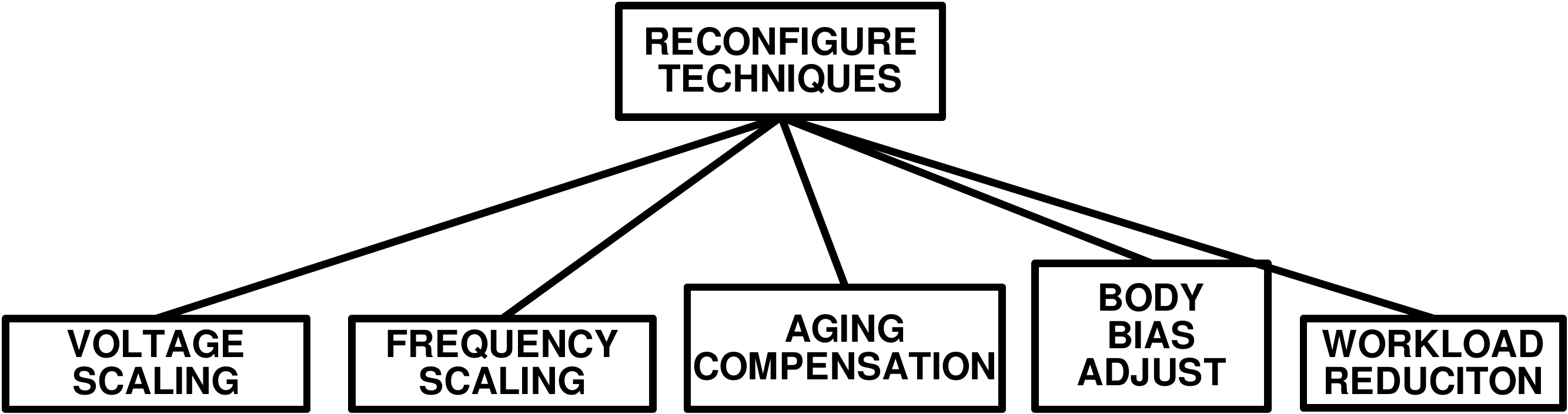}
    \caption{Reconfiguration techniques covered by the survey.}
    \label{fig:reconf_tree}
\end{figure}

This Section answers the second research question of this survey: ``What solutions the literature presents for circuit reconfiguration?''. \Cref{fig:reconf_tree} and \Cref{tab:reconf_class} present the reconfiguration techniques covered in this survey. According to the Table, voltage scaling is the most adopted reconfiguration technique, present in 50\% of the papers about reconfiguration techniques. The second most common reconfiguration technique used for aging detection is frequency scaling, which is present in 28.57\% of the papers.

%
%   FIGURE 5 
%

\tabReconfig{!ht}

%%%%%%%%%%%%%%%%%%%%%%%%%%%%%%%%%%%%%%%%%%%%%%%%%%%%%%%%%%%%%%%
%%%%%%%%%%%%%%%%%%%%%%%%%%%%%%%%%%%%%%%%%%%%%%%%%%%%%%%%%%%%%%%
%%%%%%%%%%%%%%%%%%%%%%%%%%%%%%%%%%%%%%%%%%%%%%%%%%%%%%%%%%%%%%%
%%%%%%%%%%%%%%%%%%%%%%%%%%%%%%%%%%%%%%%%%%%%%%%%%%%%%%%%%%%%%%%

%%%%%%%%%%%%%%%%%%%%%%%%%%%%%%%%%%%%%%%%%%%%%%%%%%%%%%%%%%%%%%%
%%% Literature Review
%%%%%%%%%%%%%%%%%%%%%%%%%%%%%%%%%%%%%%%%%%%%%%%%%%%%%%%%%%%%%%%
\section{Literature Review}
\label{sec:lit_review}

This Section describes the papers related to aging monitors and reconfiguration techniques covering the years from 2012 to 2019. The subsections are grouped by monitoring approaches, according to the classification proposed in \cref{sec:prop_class} for aging monitors. When a paper also presents a reconfiguration technique, it is described together, in the same paragraph.

%%%%%%%%%%%%%%%%%%%%%%%%%%%%%%%%%%%%%%%%%%%%%%%%%%%%%%%%%%%%%%%
%%%%%%%%%%%%%%%%%%%%%%%%%%%%%%%%%%%%%%%%%%%%%%%%%%%%%%%%%%%%%%%
\subsection{Temperature Monitors}

NBTI, PBTI, and HCI are aging effects accelerated by temperature increase in chips. System workload affects power dissipation, which has a direct impact on the temperature. Thus, monitoring temperature helps to deal with these effects. Some sensors use ring-oscillators to capture the aging effects caused by temperature increase. This kind of sensor provides an indirect measurement of temperature, and can be considered a temperature monitor. 

%- An On-die Digital Aging Monitor against HCI and xBTI in 16 nm Fin-FET Bulk CMOS Technology
%(Synchronous, digital, software)
% Reconf - DVS
% Reconf - DFS
Igarashi et al. \cite{igarashi2015digital} propose an aging monitor implemented with ring-oscillator (RO) to measures BTI and AC hot-carrier-injection (AC-HCI). The monitor consists of a symmetric RO (SRO) and an asymmetric RO (ASRO). ASRO is an RO composed of standard cells of different drives, while SRO is implemented only by standard cells with the same driving strength. With these two types of RO, it is possible to separate NBTI and PBTI effects for analyses by observing them under DC stress conditions. Also, the speed degradation caused by AC-HCI can be detected because unbalanced delay with a long/short transition in ASRO has high sensitivity against AC-HCI under AC stress. A dynamic voltage and frequency scaling (DVFS) technique controlled by software is used to change the supply voltage and clock activity dynamically and reconfigure the circuit. A test chip, including both SRO and ASRO using NAND2 standard cells, was implemented in a 16 nm Fin-FET bulk CMOS technology. Results show that $V_{th}$ shift due to PBTI measured from frequency degradation is 2mV, which is still 1/10 of NBTI in Fin-FET technology, and that is possible to reduce the BTI guard bands in 45\% at the nominal frequency operation. 

%Embedded Silicon Odometers for Monitoring the Aging of High-Temperature Integrated Circuits
%Estimating Circuit Aging Due to BTI and HCI Using Ring-Oscillator-Based Sensors
%Investigation of HCI effects in FinFET based Ring Oscillator Circuits and IP Blocks
%(Synchronous, digital) 
Majerus et al. \cite{majerus2017embedded} use ROs to measure changes in transistor and resistor parameters as a function of the stress caused by aging. The result is a data-driven aging model that provides information that can be used to ensure system reliability.

Sengupta and Sapatnekar \cite{sengupta2017estimating} present two methods that use sensors implemented with ROs to detect the delay shifts in circuits as a result of BTI and HCI effects. The first method uses a pre-silicon analysis of the circuit to compute calibration factors that can translate the delay shifts in the ROs with a delay estimate of 1\% of the real values. The second method uses an analysis where sensor measurements are combined with infrequent online delay measurements to reduce the circuit guard bands and allows 8\% lower delay guard banding overheads compared to the conventional methods.

Kim et al. \cite{kim2017investigation} propose a new test structure of RO that helps to measure the stress duty cycle (SDC) of the HCI. SDC is the ratio between the stress caused by the aging effect and the circuit cycle time \cite{hofmann2010highly}. The structure is composed of a NAND gate that has the function of enabling the oscillator mode and inverters. VDD and GND bias of the HCI stress inverter are set up in a complementary way during the stress, making the device not suffering from BTI aging. Also, a buffer is designed as a compensator for the signal falling by as much as $V_{th}$. Results demonstrated that HCI SDC increases with frequency, but the maximum duty cycle was much less than 2\%. 

%- Performance optimization for on-chip sensors to detect recycled ICs
%- Temperature and Voltage Measurement for Field Test Using an Aging-Tolerant Monitor
%(Synchronous, digital) 
Shakya et al. \cite{shakya2015performance} propose an NBTI sensor that uses two ROs, one without circuit influence used as a reference and other to evaluate the degradation circuit under stress, and compare the output of both. This approach gives the manufacturer the exact control over the yield and accuracy of the sensors, which not occurs with ad hoc approaches that determine parameters such as the decision threshold. 

Miyake et al. \cite{miyake2016temperature} propose an aging-tolerant monitor that analyzes frequencies of more than one RO. Thus, it is possible to derive the values for temperature and voltage from the frequencies using multiple regression analysis. Besides, three techniques to select the RO types are proposed to improve the accuracy of the measurement. The method was validated with simulations in 180 nm, 90 nm, and 45 nm CMOS technologies. In the 180 nm technology, temperature accuracy is about 0.99$^{\circ}$C, and voltage accuracy is about 4.17 mV. Also, the authors fabricated test chips with 180 nm CMOS technology to confirm its feasibility.

%- On-Chip Aging Compensation for Output Driver
%(Synchronous, digital)
% Reconf - Aging comp
Kumar \cite{kumar2014chip} presents an aging compensation technique for a CMOS transistor output driver. It contains an aging compensation cell which monitors the degradation in the ON current ($I_{ON}$) of output driver due to the HCI and BTI effect. Based on this degradation, the aging compensation cell generates compensation codes for PMOS and NMOS transistors drivers. These aging compensation codes turn on devices in parallel to the main driver, increasing the drive strength of output, which compensates the degradation in NMOS and PMOS driver due to HCI and BTI effects. The design was implemented in 40 nm CMOS process by using 1.8V thick-oxide devices. Results show that by using the proposed aging compensation technique, the impact of aging on output driver reduces by 70\% after ten years of operation.

%- Accessing On-Chip Temperature Health Monitors Using the IEEE 1687 Standard
%  (Synchronous, analog and digital)
Ali et al. \cite{ali2016accessing} use IJTAG to manage temperature health monitors on the chip. The temperature health monitors are based on a Wheatstone bridge, which is composed of four resistors, one operational amplifier, one filter, and one analog to digital converter. Results show that the proposed solution can be used to manage and control instruments to ensure the reliable operation of the chip over its lifetime. According to the authors, the proposed method can reduce the overall time spent on the test, once there is no off-chip interface, and the system clock can be used instead of the test clock.

Rathore et al. \cite{rathore2019lifeguard} propose to use temperature and NBTI sensors to implement a task mapping strategy to manycore systems, called LifeGuard. LifeGuard considers performance and aging as parameters to perform task mapping, and is based on reinforcement learning. At each tile of the network-on-chip (NoC), an NBTI and a thermal sensor are added and used to perform the task mapping. As a benefit, LifeGuard prevents the rapid aging of cores that map a more significant number of tasks. Also, it improves the aggregate safe operating frequency of the system. Experimental results using a 256-core system showed that LifeGuard improved the health of the cores for 57\% when compared to the HiMap strategy \cite{rathore2018himap}, and 74\% when compared to the Hayat \cite{gnad2015hayat}. Also, LifeGuard shows a better system performance.

%%%%%%%%%%%%%%%%%%%%%%%%%%%%%%%%%%%%%%%%%%%%%%%%%%%%%%%%%%%%%%%
%%%%%%%%%%%%%%%%%%%%%%%%%%%%%%%%%%%%%%%%%%%%%%%%%%%%%%%%%%%%%%%
\subsection{Critical Path Monitors}

Contemporary literature presents monitors that capture timing errors on the critical paths of circuits (Definition \ref{def_cp}). This technique contains extra components that capture the memory register output of the critical paths and its processed data. A comparison between these data is executed to identify if a timing error occurred.

%- A BIST Approach for Counterfeit Circuit Detection based on NBTI Degradation
%  (Synchronous, digital)
% Figure \ref{fig:cda} shows the circuit architecture.
Savanur et al. \cite{savanur2015bist} present a BIST (Definition \ref{def_bist}) approach to detect aging effects on the circuit during test mode.  The BIST circuitry uses two identical buffer chains and a logic block to compare the output of these chains. If there is a difference between the outputs of the chains, an error caused by aging is detected. The approach needs extra components, which are a D flip-flop, a NAND gate, an AND gate, and two buffers. HSPICE simulations on 45 nm and 65 nm were performed to extract results. This paper focuses only on the NBTI aging factor, and results show that the solution can detect minimal stress levels in the presence of process variations and that the aging detection depends on the time that a path of the circuit keeps at logic level zero.

%- A Cost-efficient Delay-Fault Monitor
%  (Synchronous, digital)
%, as showed in Figure \ref{fig:ppc}
Sai et al. \cite{sai2017cost} present a Parity Check Circuit (PPC) for monitoring delay-faults and compare it with the Canary flip-flop approach \cite{fuketa2012adaptive}. Unlike the Canary flip-flop approach, PPC can monitor more than one logic path simultaneously, which allows reducing the number of sensors. PPC is implemented using a multiple-input XOR gate, a delay element (DE), a matched delay element (MD), a flip-flop, a shadow flip-flop, and a 2-input XOR gate. The multiple-input XOR gate is responsible for computing the data parity, and the MD is responsible for adding a delay to the clock signal to compensate delay produced by the multiple-input XOR gate. The PPC was validated in a 32 bit MIPS processor using a 65 nm technology. Results indicate that the use of the circuit reduces area overhead by 66\% and power by 33\% when compared to the Canary flip-flop approach.

%- Multi-Path Aging Sensor for Cost-Efficient Delay Fault Prediction
%(Synchronous, digital) 
Sai et al. \cite{sai2018multi} propose a metastability-free aging sensor called Differential Multiple Error Detection Sensor (DMEDS). The sensor monitors multiple paths concurrently. It is composed of a Multiple Detection Unit (MDU) and a stability checker, which allows monitoring two or more critical paths simultaneously. Any transition in the data active the MDU output signal that is captured by the stability checker, signaling a delay fault. The stability checker checks the stability of the delayed signal while the clock is high.  DMEDS was designed at transistor level using a 32 nm technology and applied to a 32-bit MIPS processor to monitor ten paths concurrently. Results show that using DMEDS for monitoring ten paths can save 197.1\% and 97.1\% in area overhead when compared to Razor \cite{ernst2003razor} and Canary \cite{fuketa2012adaptive}, respectively.

%- NBTI-Aware Design of Integrated Circuits: A Hardware-Based Approach for Increasing Circuits’ Life Time
%(Synchronous, digital)
% Reconf - DVS
Copetti et al. \cite{copetti2016nbti} propose a hardware-based technique able to increase ICs lifetime. The technique is based on a sensor able to monitor IC aging and to adjust its power supply voltage to minimize NBTI effects, increasing the circuit lifetime. The approach is composed of: (i) an aging sensor, which contains a delay element and the stability checker; (ii) an actuator, which contains a counter and a decoder block; and (iii) a flip-flop inserted at the critical path output. The flip-flop output of the critical path passes through the delay element and the stability checker while the clock signal is high, analyzing data transitions. If a transition while the clock is high occurs, it means that the delay of the circuit increased, changing the output of the stability checker and indicating a timing violation. Also, the actuator receives the signal from the stability checker and increases its counter, allowing the decoder to adjust the power supply. Experimental results obtained by simulations demonstrate that the technique increases the circuit lifetime by 150\%.

%- Online Self Adjusting Progressive Age Monitoring of Timing Variations
%(Synchronous, digital)
Sadeghi-Kohan et al. \cite{sadeghi2015online} propose a self-adjusting age monitoring method to pipeline circuits, which allows detecting progressive changes in the timing of a circuit. The output of the critical paths are captured using an age monitoring clock (that occurs before the system clock), and this captured data is compared with the same output but captured at the rising edge of the system clock. The circuit used to adjust the age monitoring clock has an age indicator counter that counts RO pulses to adjust the clock phase and is initialized with the core process characteristic. As the core ages, the age indicator counter is incremented, causing a more extended clock phase shift, and shorter slack time. The monitors are designed targeting low hardware overhead and accuracy in reported timing changes. 

In another work, Sadeghi-Kohan et al. \cite{kamalself} use a similar strategy to monitor paths of a circuit and to detect its continuous age growth. This approach can provide the aging rate and the aging state of the circuit. The proposed strategy uses a clock generator to feed the register responsible for capturing the data before the system clock. Results show an area overhead of 2.13\%, a power overhead of 0.69\%, and a low-performance overhead.   
%- A Scan-Based On-Line Aging Monitoring Scheme
%(Synchronous, digital) 
Yi et al. \cite{yi2014scan} present a scan-based on-line monitoring that monitors aging during system operation and gives an alarm if the system detects aging effects. This work inserts an extra scan chain that captures early the functional data (at the opposite edge of the system clock) within a given guard-band interval during system operation. After that, the extra scan chain output is compared to the original scan chain output, which allows detecting violations. The scan-chain scheme contains two scan-chains, one with conventional scan cells (SC scan-chain) and the other one with the early capture scan cells (ECSC scan-chain). The SC flip-flops capture the data at the clock rising edge, while the ECSC flip-flops capture the data at the clock falling edge. The modified scan chain uses an XOR gate to compare the captured data. 

%- On Diagnosing the Aging Level of Automotive Semiconductor Devices
%(Synchronous, digital) 
% Figure \ref{fig:peff} shows the PEFF, 
Jung \cite{jung2017diagnosing} presents an aging level estimating flip-flop that exploits the frequency guard band of a device to estimate the aging level with a small power overhead, called performance estimation flip-flop (PEFF). The PEFF has five elements: i) a scan flip-flop; ii) a shadow latch; iii) a sampling time indicator; iv) a logic block to controls the input of the performance result cell (PERC); v) the PERC. The shadow-latch is used to sample the data earlier than the functional flip-flop. Results show a reduction in monitoring time, and the power consumption is reduced by 50\% when compared to the Yi et al. \cite{yi2014scan}.

%- Error Prediction and Detection Methodologies for Reliable Circuit Operation under NBTI
%(Synchronous, digital) 
Vazquez-Hernandez \cite{vazquez2014error} proposes a solution for error prediction using an aging sensor based on the Error-Detection Sequential (EDS) circuit \cite{bowman201145}. The EDS has a decoder module to monitors the critical paths. When one of the paths is activated, the decoder active the EDS to allow detect errors. The methodology for path selection uses statistical static timing analysis. Results show that the EDS can reduce power overhead form 102\% to 6\% and area overhead from 69\% to 22\% when compared to \cite{agarwal2008optimized} and \cite{yi2014scan}, considering the circuit characteristics as the number of gates and buffers. 

%- Monitoring Reliability in Embedded Processors – A Multi-layer View
%(Synchronous, digital) 
Chandra \cite{chandra2014monitoring} proposes the SlackProbe monitor. This approach inserts timing monitors at endpoints and intermediate nodes of the circuit paths. If a monitor is inserted at an intermediate node, an AND gate is used as delay matching, and a transition detector is connected to the intermediate node with a minimum size inverter. If a signal transition at the intermediate node occurs, it arrives at the transition detector through the delay chain, and the signal is compared with the incoming clock edge. If the transition is close to its required arrival time, a corresponding signal transition arrives at the transition detector input after the clock edge. This transition triggers the transition detector and flags a signal indicating a delay failure. The monitor inserted at the intermediate node is capable of monitoring the delay of all critical paths passing through it, and its output can be used for mitigating failures due to aging (based on hardware or software). The results show that SlackProbe can achieve up to 16x reduction in the total number of monitors.

%- MTTF-aware Design Methodology of Error Prediction Based Adaptively Voltage-scaled Circuits
%(Synchronous, digital)
%¨Reconf - DVS
Masuda and Hashimoto \cite{masuda2018mttf} propose an error prediction adaptive voltage scaling (EP-AVS) and a mean time to failure aware (MTTF-aware) design methodology for EP-AVS circuits. The EP-AVS has a main circuit plus a timing error predictive flip-flop (TEP-FF) and a voltage control unit. The TEP-FF has a flip-flop, delay buffers, and a comparator implemented with an XOR gate. Also, TEP-FF works with the main flip-flop. When the timing margin is gradually decreasing, a timing error occurs at the TEP-FF before the main flip-flop captures a wrong value due to the delay buffer. This wrong value produces an error prediction signal, which allows the voltage control logic to provide a higher supply voltage and reduce the circuit delay. Evaluation results show that the proposed EP-AVS design methodology achieves a 20.8\% voltage reduction while satisfying the target MTTF.

%- Workload-aware Static Aging Monitoring and Mitigation of Timing-critical Flip-flops
%(Synchronous, digital, software)
% Reconf - workload
Vijayan et al. \cite{vijayan2017workload} propose an aging monitor based on hardware and software. The system is composed of representative flip-flops (RFF) that are selected in an offline phase and connected to the monitoring hardware. The aging effect consists of two phases: i) stress phase, where the transistor is under the aging effect; ii) recovery phase, where the transistor is recovering from the stress phase. A switching event in the RFF corresponds to the recovery phase of the corresponding flip-flop group, which is captured to report the recovery event to the software. The representative flip-flops are observed by a switching-event detector to perform the capture operation, which is composed of an XOR gate and a shadow flip-flop that generates a pulse at its output when a logic transition occurs in the corresponding flip-flop. The output of the switching-event detector is encoded using a priority encoder. In a determined clock cycle, the output of the priority encoder indicates the index of the flip-flop that switches its state in that particular clock cycle. If two representative flip-flops change their states at the same time, the priority encoder ensures a valid output. The critical-flag register (CFF) keeps the criticality word to represent the recovery/aging state of the corresponding representative flip-flop. This paper uses a software subroutine to mitigate aging by inserting NOPs instructions, which allows a relaxation on BTI stress, reducing the workload. Results show that area and power overheads imposed by the monitoring hardware are less than 0.25\% for a Leon3 processor and Fabscalar processor.

%- Digital Circuits Reliability with In-Situ Monitors in 28nm Fully Depleted SOI
%(Synchronous, analog and digital) 
Saliva et al. \cite{saliva2015digital} propose monitors based on delay elements called pre-error flip-flops. The approaches are composed of a shadow flip-flop that stores delayed data, and is works in parallel to the regular flip-flop. The approach compares the two flip-flop outputs, and a pre-error signal is generated to predict the occurrence of timing errors. Each monitor uses different delay approaches. The first approach is the buffer delay, where buffers produce the delay. The second is the passive delay, where a resistor generates the delay. The last is the master delay, where master-slave latches replace the regular flip-flop, and the slave latch outputs feed the shadow flip-flop to generate the delay. Results show that the detection window of the in-situ monitors with passive delay is less deviant than the buffer and master delay ones with $V_{dd}$ decrease. However, using a passive element in a digital circuit is not common. The in-situ monitor with buffer delay is a better choice because it uses standard cells in its implementation. 

Di Natale et al. \cite{di2019hidden} propose a hidden-delay-fault sensor that can be used to detect small delay faults. The sensor allows the circuit to operate at the nominal frequency, and it is inserted in a critical path. The monitor works sampling a signal in both clock edges. After that, the monitor compares the two samples using an XOR gate and stores the result on the next falling edge of the clock suing a shadow flip-flop. Result extraction is performed using a classic scan chain. The authors propose that the sensor can be used during the lifetime of the circuit to identify timing violations in short paths caused by aging. Also, the authors mention that it is possible to use the sensor combined with reconfiguration techniques such as DVFS. The paper does not present results.

Wang et al. \cite{wang2019aging} uses a timing margin detector (TMD) to monitor aging and capture delay behavior. Also, the output detector is used in a machine learning engine based on a support vector machine (SVM) to predict aging. The TMD is used to capture late transitions. It is composed of two D flip-flops, and by an OR gate at the flip-flops output. Results show that it is possible to obtain a 97.40\% of accuracy in aging prediction, with 4.14\% area overhead on average.

%%%%%
%SEU%
%%%%%

%- A Low-Overhead Integrated Aging and SEU Sensor
%(Synchronous, digital)
Rohbani and Miremadi \cite{rohbani2018low} propose an aging sensor combined with the flip-flops of the design that monitors the critical path output before the rising edge of the clock signal. This signal follows the system clock by an adjusted delay of about 10\% to 20\% of the clock period. When the clock is at logic level one, the sensor is activated. Any change in the input signal during the period where the sensor clock is at logic level one and the system clock is at level logic level zero represents a delay extension of the critical path due to aging effects. Results show that the precision of the proposed sensor is about 2.7 higher, with almost 33\% less area overhead compared with state-of-the-art aging sensors. Furthermore, the presented sensor can detect and correct 50\% of the Single Event Upsets (SEUs), which lead to a bit-flip in the flip-flops. Besides, the SEU detection circuitry can reduce Bias Temperature Instability (BTI) by balancing the duty cycle of the flip-flop with negligible extra overhead.

%%%%%%%
%DELAY%
%%%%%%%

%- Aging-Aware Power or Frequency Tuning With Predictive Fault Detection
%- Aging-Aware Dynamic Voltage or Frequency Scaling
%(Synchronous, digital)
% Reconf - DVS
% Reconf - DFS
Pachito et al. \cite{pachito2012aging} propose an aging-aware power supply or frequency reconfiguration approach that uses global and local sensors. Global sensors perform periodic or on-demand delay monitoring, while local sensors predict errors locally. Both allow adjusting frequency or power supply voltage. Results show that performance and power can be improved by, respectively, increasing the frequency and reducing the voltage while still preventing errors. In other work, Semião et al. \cite{semiao2014aging, semiao2014performance} propose improvements in the global and local sensors cited previously. The global sensor was improved to detect Negative-bias temperature instability (NBTI) and  Positive-bias temperature instability (PBTI), while the local sensor was improved to tolerate delay-faults.

%- Aging-Aware Adaptive Voltage Scaling in 22nm High-K/Metal-Gate Tri-Gate CMOS
%(Synchronous, digital)
% Reconf - DVS
Cho et al. \cite{cho2015aging} examine the effectiveness of the aging-aware Adaptive Voltage Scaling (AVS) for logic circuit blocks using a Tunable Replica Circuit (TRC) aging monitors. The TRC is calibrated off-line, based on the critical paths and on-line monitoring of the operational conditions of the circuit, as temperature variations. The Power Management Unit (PMU) communicates with the TRC periodically. The PMU tunes the Voltage Regulator Module (VRM)  when the TRC detects an aging delay degradation until the TRC detects the correct behavior based on the clock. Simulation results in a 22 nm High-K/Metal-Gate Tri-Gate CMOS process show a 7\% power reduction with the removal of the guard-bands of a conventional fixed $V_{cc}$.

%- Design a Delay Amplified Digital Aging Sensor Circuit in 65nm CMOS
%(Synchronous, digital) 
Ding et al. \cite{ding2016design} propose a delay amplified digital (DAD) aging sensor circuit composed of a delay sensor and a signal amplification circuit. It uses a reference delay circuit designed according to the monitored combinational logic circuit. The delay sensor is used to detect aging effects, while a timing multiplier circuit eliminates the effects on the environment, improving the aging sensor data accuracy. A digital sample module uses nine T flip-flops to count the number of falling-edge during the amplified enable pulse. Using the parameters of TSMC 65 nm CMOS technology, the DAD sensor circuit is designed and simulated using SPECTRE.

%- Robust and In-Situ Self-Testing Technique for Monitoring Device Aging Effects in Pipeline Circuits
%(Synchronous, digital)
% Reconf - DVS
Li and Seok \cite{li2014robust} propose a technique to pipeline circuits that enables accurate of aging monitoring even under environmental variations. The technique scales the supply voltage for a temperature-insensitive delay and reconfigures the target paths into ring oscillators. The oscillation periods are measured and compared to pre-aging measurements to estimate the delay degradation caused by aging. Also, the technique presents a new register implementation, that has an area overhead of 12 transistors when compared to a standard flip-flop, and a relatively low delay overhead, once it adds a small amount of load between the path from input to output. The technique adds a feedback network between the input and output registers of target paths, which can present a small portion of the total oscillation period, making little impact on monitoring accuracy. The area overhead of feedback network can be minimized by sharing feedback paths among multiple target paths, as like the counter. The counter is used to measure the periods of the ring oscillator operation. Results show that the technique achieves highly-accurate monitoring with an error of 15.5\% across the temperature variations in self-test phases from 0$^{\circ}$C to 80$^{\circ}$C, exhibiting more than 30 times improvement in accuracy as compared to the conventional technique operating at the nominal supply voltage.

%- On-chip Aging Prediction Circuit in Nanometer Digital Circuits
%(Synchronous, digital) 
Jang et al. \cite{jang2014chip} propose a aging sensor that detects failures caused by BTI and HCI. This aging sensor is based on timing warning windows to detect a guardband violation of sequential circuits and generates a warning right before circuit failures occur. It monitors the moment when the critical path delays of the logic exceed a standard value, which guarantees a correct circuit operation. The aging sensor is composed by: i) a guardband generator; ii) a path delay monitor; iii) a hold circuit; iv) a signal that controls the aging sensor. The circuit was implemented in 110 nm, and results show that the aging sensor achieves a good aging failure prediction with low overhead.

\Cref{tab:path_delay_summary} summarizes the reviewed path delay monitors. Most of the monitors use delay elements and comparison mechanism to detect timing errors. Also, most of the approaches monitor just the critical paths without monitoring other system elements.

\tabCriticalPath{!ht}

%%%%%%%%%%%%%%%%%%%%%%%%%%%%%%%%%%%%%%%%%%%%%%%%%%%%%%%%%%%%%%%
%%%%%%%%%%%%%%%%%%%%%%%%%%%%%%%%%%%%%%%%%%%%%%%%%%%%%%%%%%%%%%%
\subsection{Clock Frequency Monitors}

%- Aging Adaption in Integrated Circuits Using a Novel Built-In Sensor
%  (Synchronous, digital)
% Reconf - DFS
Aging can affect clock frequency and decreases system performance. Thus, clock frequency monitoring is a way to detect aging in a design.

Wang et al. \cite{wang2015aging} present a sensor for reliability analysis of digital circuits using standard-cells called Radic. The Authors also propose a low-cost built-in aging adaption system based on the Radic sensor to perform in-field aging adaption. Radic allows frequency, aging, and metastability measurements. Also, the sensor is designed to obtain the frequency difference between the waveform under test and a reference frequency by measuring how much clock cycles the wave under test is faster or slower than the reference frequency. A stable external source such as automatic test equipment, or a waveform generator, or an internal source such as a phase-locked loop (PLL) can generate the reference frequency. An m-bit timer stores the length of the measurement window by counting the number of clock cycles of the reference frequency. The Radic-based aging monitor system is inserted into a Freescale IP and an ITC'99 b19 benchmark. Results show that the system reduces the fixed aging guardband by 80\%. A comparison between the original design and the design with the proposed adaption system shows an area reduction of about 1.02\% to 3.16\% in most cases. Power is also reduced, as the design can be synthesized using smaller drive strength. 

%Thermally-aware Sensor Allocation for Real-time Monitoring and Mitigation of FEOL Aging in System-on-Chip (SoC) Applications
%(Synchronous, digital) 
Küflüoglu et al. \cite{kufluoglu2017thermally} propose an aging sensor based on RO that uses two identical aging paths. Both paths can be either equally sensitive to BTI or modulated to be more sensitive to aging effects. Using DC biased with opposite polarity inputs, both paths have the PBTI and NBTI effects stressed alternatively at every other stage.  Unlike a conventional RO based aging sensor, a  control loop logic links all stressed devices into one measuring RO loop which its output is the frequency degraded due to the aging. Also, the control loop logic links all non-stressed devices are linked into a second RO loop, wich the output is the reference frequency. The frequency delta between aging frequency and reference frequency is used to monitors aging.

%%%%%%%%%%%%%%%%%%%%%%%%%%%%%%%%%%%%%%%%%%%%%%%%%%%%%%%%%%%%%%%
%%%%%%%%%%%%%%%%%%%%%%%%%%%%%%%%%%%%%%%%%%%%%%%%%%%%%%%%%%%%%%%
\subsection{Circuit State and Workload Monitors}

%- Fine-Grained Aging Prediction Based on the Monitoring of runtime Stress Using DfT Infrastructure
%(Synchronous, digital, software) 

It is possible to detect aging by monitoring the whole system instead of just critical paths. The circuit state and its workload are metrics that can be monitored and provide insightful information. For example, they can indicate stress levels that will increase the aging effects impact.

Koneru et al. \cite{koneru2015fine} reuse the design for testability (DfT -- Definition \ref{def_dft}) infrastructure to perform a fine-grain workload-induced stress monitoring for accurate aging prediction. A multiple-input signature register (MISR) is used to capture the workload effect on the circuit. An aging prediction software, based on support vector machine (SVM) learning technique, performs aging mitigation. The DfT controller, implemented as a finite-state machine (FSM), periodically switches the circuit into scan mode to capture the circuit state. After capturing the state, the contents of the scan chains are then shifted out to the MISR. The scan-chains are modified to keep its value during the aging monitoring phase, which overwrites the state of the flip-flops and not allow the circuit to return to normal operation until completing the shift to MISR. The Authors conducted experiments on two open-source processor benchmarks, namely OpenRISC 1200 and Leon3, and on four ISCAS'89 benchmarks, to evaluate the accuracy of the proposed technique. Simulation results show that the proposed approach can accurately predict workload-induced aging trends. In a similar work, Firouzi et al. \cite{firouzi2015re} reuse the BIST structure to predict the fine-grained circuit-delay degradation with minimal area and performance overhead and high accuracy.

%- A Self-Adaptive System Architecture to Address Transistor Aging
%(Synchronous, digital, software) 
% Reconf - DVS
% Reconf - DFS
Khan and Kundu \cite{khan2009self} propose a system-level reliability management scheme (SRM) that dynamically adjusts the operating frequency and supply voltage according to the system aging. The proposal allows continuous runtime adjustments based on parameters such as actual room temperature and power supply tolerance. The SRM communicates with the voltage and frequency control registers to enable frequency and voltage reconfiguration. It is implemented in software and assumes a Virtual Machine Monitor (VMM) running underneath the OS software stack, which is primarily used to enter and exit the SRM. The SRM software enables carefully crafted functional stress tests or built-in self-test control to identify degradation at a component granularity and provides adjustments for sustained performance levels at the target reliability. The software allows the system to adapt to the aging effects and invokes aging device management at determined periods. The results show that the device can operate near peak frequency throughout product life. Also, the approach ensures protection against failure due to insufficient lifetime guardband and no system downtime or change.

%- Design of Reliable SoCs With BIST Hardware and Machine Learning
%(Synchronous, digital, software)
% Reconf - DVS
% Reconf - DFS
% Reconf - BBA
Sadi et al. \cite{sadi2017design} presents a framework for designing lifetime-reliable system-on-chip (SoC) with reconfiguration capability to deal with aging effects. The proposed flow uses a BIST (Definition \ref{def_bist}), and a machine learning linear regression predictor software to activate aging countermeasures. The aging status of the chip is monitored at regular intervals by the BIST hardware. Based on the observations, proactive adaptation methods are taken to counteract the reliability degradation effect. The framework allows testing patterns from SoC's existing BIST hardware, collect the response, and tune the linear regression software. A gate-overlap and path-delay-aware algorithm selects a minimum set of patterns, which activate the target paths used as features of the linear regression predictor. The seeds of the selected patterns are stored in on-chip memory and applied at the BIST hardware at multiple test clock frequencies when required. The corresponding responses of these patterns are collected in a separate response storage flip-flop chain. The software-implemented machine learning classifier is trained with the collected multiple-frequency responses, and the trained predictor accurately predicts the state of aging degradation at runtime. The paths to be monitored by the BIST hardware are selected at the design time based on timing analysis. The adaptive methods used in this approach are frequency scaling, voltage scaling, and adaptive body biasing. Simulation results show that the proposed technique allows accurate and fine-grained in-field aging prediction, with a precision that can reach 94\%. 

%- On-Line Prediction of NBTI-induced Aging Rates
%(Synchronous, digital, software) 
Baranowski et al. \cite{baranowski2015line} present a method for aging rate prediction, which is based on workload monitoring and linear regression machine learning technique. The monitoring technique enables the on-line prediction of the degradation rate caused by the currently running application. The degradation rate monitoring system is composed of a workload monitor and a temperature sensor. The linear regression machine learning technique is used to find the representative critical gates that are monitored. Results show that this method delivers sufficient accuracy at an area overhead of 4.2\%, which decreases with the size of the monitored circuit.

%%%%%%%%%%%%%%%%%%%%%%%%%%%%%%%%%%%%%%%%%%%%%%%%%%%%%%%%%%%%%%%
%%%%%%%%%%%%%%%%%%%%%%%%%%%%%%%%%%%%%%%%%%%%%%%%%%%%%%%%%%%%%%%
\subsection{Voltage Monitors}

%- NBTI Detection Methodology for Building tolerance with respect to NBTI effects Employing Adaptive Body Bias
%(Synchronous, digital and analog) 
% Reconf - BBA
Circuit voltage can also be use to monitor aging effects. Similar to what happens in clock frequency variations, as the circuit ages, negative effects will be observed, such as timing errors and path delay extension.

Narang and Srivastava \cite{narang2015nbti} propose an approach that uses an inverter chain and counter-based technique to detect the variation in the threshold voltage. A voltage sensing methodology is used to monitor the circuit's voltage variation. This variation can be fed to an Adaptive Body Bias (ABB) circuit to mitigate the effects of NBTI. The approach uses a counter to determine the total path delay. The path delay feeds an inverter connected to an amplifier. The amplifier output pass trough a voltage to current converter, which feeds a logarithmic amplifier. The output of this logarithmic amplifier feds an exponential function generator that passes through a subtractor circuit that is used to detect a change in the threshold voltage due to the NBTI effect. The circuit was validated in a set of circuits such as 32-bit OR gate, 64-input OR gate, and 32-bit comparator in 32 nm technology. The simulation results show that this methodology is efficient as it reduces the delay to a large extent with minimal increase in power.

Xiaojin et al. \cite{li2018linear} propose a digital on-chip detector that uses the circuit output voltage phase to detect aging. The approach consists of duplicating the circuit, generating two circuits: a reference circuit and an aging stressed circuit. The output of both is compared using an XNOR gate. If a pulse occurs in the XNOR output, aging is detected. Also, the XNOR output pass trough time to digital converter to facilitate the circuit state analyses and verification. The authors claim that the aging detecting circuit can be applied in adaptive systems to mitigate the aging, but do not present a solution in this work. The validation was made by a chip implementation and by simulation. The chip demonstrates results close to the simulated regarding aging time.

%%%%%%%%%%%%%%%%%%%%%%%%%%%%%%%%%%%%%%%%%%%%%%%%%%%%%%%%%%%%%%%
%%%%%%%%%%%%%%%%%%%%%%%%%%%%%%%%%%%%%%%%%%%%%%%%%%%%%%%%%%%%%%%
%%%%%%%%%%%%%%%%%%%%%%%%%%%%%%%%%%%%%%%%%%%%%%%%%%%%%%%%%%%%%%%

\section{Conclusion} 
\label{sec:conc}

This Section concludes the aging monitors and reconfiguration techniques survey, providing insights for future research and development. \Cref{tab:class} presents the implementation methods.

\tabConclusion{t}

Most of the literature contributions are in the digital area, specifically using hardware solutions for monitoring aging in circuits. Few works use software approaches for aging monitoring. Most software applications act on reconfiguring the circuit after aging detection. We observed works using learning methods (software) for taking proactive actions, detecting events related to aging before they occur. These learning-based methods can point out a promising way to detect the effects of aging, as long as they are associated with a rich set of hardware monitors, such as temperature monitors. Also, few contributions combine analog and digital techniques to perform aging monitoring.

As mentioned, the most common monitor used for aging detection is the timing error monitor. However, this technique is a function of the paths chosen to be monitored during design time. If this choice is executed incorrectly and the set of critical paths does not represent the critical paths, the aging counter-effects solutions are inefficient. With the CMOS scaling, the number of critical paths increases, which increases the number of components to monitor the aging effects. Thus, it is likely that the use of solutions focused on timing errors and path monitoring will decrease in the future or be limited to older technologies nodes. 

We believe that solutions based on system-wide monitoring parameters such as temperature, voltage, and frequency will be more prevalent once they do not depend on a specific parameter.

Summarizing, current techniques heavily rely on timing error monitors for detecting aging and use voltage scaling to compensate the effects in integrated circuits. We believe that new solutions are required due to the fact that as technology nodes advance, it becomes harder to identify critical paths of a circuit. This limitation not only makes the task of defining where to place timing error monitors more challenging, but can also require an increase in the number of monitors, inducing larger area and power overheads. Furthermore, with the availability of different sensors, such as frequency and temperature, in industrial SoCs, we believe there is a gap to be filled in aging detection algorithms. Especially with emerging artificial intelligence capabilities, algorithms can analyze the data of the available sensors and configure the SoC to counter aging effects.

%#===========================================================
%# SECTION: REFERENCES 
%#===========================================================
%\bibliographystyle{IEEEtran}
%\bibliography{sample}

\begin{thebibliography}{1}

\bibitem{kim2020aging}
H.~Kim, J.~Kim, H.~Amrouch, J.~Henkel, A.~Gerstlauer, K.~Choi, and H.~Park,
  ``Aging compensation with dynamic computation approximation,'' \emph{IEEE
  Transactions on Circuits and Systems I: Regular Papers}, 2020.

\bibitem{sai2020cost}
G.~Sai, M.~Zwolinski, and B.~Halak, ``A cost-efficient aging sensor based on
  multiple paths delay fault,'' \emph{Ageing of Integrated Circuits: Causes,
  Effects and Mitigation Techniques}, p. 211, 2020.

\bibitem{sahoo2020novel}
S.~R. Sahoo and K.~Mahapatra, ``A novel area efficient on-chip ro-sensor for
  recycled ic detection,'' \emph{Integration}, vol.~70, pp. 138--150, 2020.

\bibitem{lu2009statistical}
Y.~Lu, L.~Shang, H.~Zhou, H.~Zhu, F.~Yang, and X.~Zeng, ``{Statistical
  Reliability Analysis under Process Variation and Aging Effects},'' in
  \emph{Design Automation Conference (DAC)}.\hskip 1em plus 0.5em minus
  0.4em\relax IEEE, 2009, pp. 514--519.

\bibitem{agarwal2008optimized}
M.~Agarwal, V.~Balakrishnan, A.~Bhuyan, K.~Kim, B.~C. Paul, W.~Wang, B.~Yang,
  Y.~Cao, and S.~Mitra, ``{Optimized Circuit Failure Prediction for Aging:
  Practicality and Promise},'' in \emph{International Test Conference
  (ITC)}.\hskip 1em plus 0.5em minus 0.4em\relax IEEE, 2008, pp. 1--10.

\bibitem{baker2019cmos}
R.~J. Baker, \emph{{CMOS}: {C}ircuit {D}esign, {L}ayout, and
  {S}imulation}.\hskip 1em plus 0.5em minus 0.4em\relax Wiley-IEEE press, 2019.

\bibitem{rahimipour2012survey}
S.~{Rahimipour}, W.~N. {Flayyih}, I.~{El-Azhary}, S.~{Shafie}, and F.~Z.
  {Rokhani}, ``{A survey of On-chip Monitors},'' in \emph{International
  Conference on Circuits and Systems (ICCAS)}.\hskip 1em plus 0.5em minus
  0.4em\relax IEEE, 2012, pp. 243--248.

\bibitem{khoshavi2017contemporary}
N.~Khoshavi, R.~A. Ashraf, R.~F. DeMara, S.~Kiamehr, F.~Oboril, and M.~B.
  Tahoori, ``{Contemporary CMOS Aging Mitigation Techniques: Survey, Taxonomy,
  and Methods},'' \emph{Integration, the VLSI Journal}, vol.~59, pp. 10--22,
  2017.

\bibitem{kochte2017self}
M.~A. Kochte and H.-J. Wunderlich, ``{Self-Test and Diagnosis for Self-Aware
  Systems},'' \emph{IEEE Design \& Test}, vol.~35, no.~7, pp. 7--18, 2018.

\bibitem{jagirdar2007efficient}
A.~Jagirdar, R.~Oliveira, and T.~J. Chakraborty, ``{Efficient flip-flop designs
  for {SET/SEU} mitigation with tolerance to crosstalk induced signal
  delays},'' in \emph{IEEE Silicon Errors Logic System Effects (SELSE)}, 2007,
  pp. 1--6.

\bibitem{schroder2003negative}
D.~K. Schroder and J.~A. Babcock, ``{Negative bias temperature instability:
  Road to cross in deep submicron silicon semiconductor manufacturing},''
  \emph{Journal of applied Physics}, vol.~94, no.~1, pp. 1--18, 2003.

\bibitem{maricau2011transistor}
E.~Maricau and G.~Gielen, ``{Transistor aging-induced degradation of analog
  circuits: Impact analysis and design guidelines},'' in \emph{European
  Solid-State Device Research Conference (ESSCIRC)}.\hskip 1em plus 0.5em minus
  0.4em\relax IEEE, 2011, pp. 243--246.

\bibitem{novak2015transistor}
S.~Novak, C.~Parker, D.~Becher, M.~Liu, M.~Agostinelli, M.~Chahal, P.~Packan,
  P.~Nayak, S.~Ramey, and S.~Natarajan, ``Transistor aging and reliability in
  14nm tri-gate technology,'' in \emph{2015 IEEE International Reliability
  Physics Symposium}.\hskip 1em plus 0.5em minus 0.4em\relax IEEE, 2015, pp.
  2F--2.

\bibitem{Book3}
M.~Abramovici, M.~A. Breuer, and A.~D. Friedman, \emph{{Digital Systems Testing
  And Testable Design}}, 1st~ed.\hskip 1em plus 0.5em minus 0.4em\relax
  Wiley-IEEE Press, 1994.

\bibitem{ernst2003razor}
D.~Ernst, N.~S. Kim, S.~Das, S.~Pant, R.~Rao, T.~Pham, C.~Ziesler, D.~Blaauw,
  T.~Austin, K.~Flautner \emph{et~al.}, ``{Razor: A Low-power Pipeline Based on
  Circuit-level Timing Speculation},'' in \emph{International Symposium on
  Microarchitecture (MICRO)}.\hskip 1em plus 0.5em minus 0.4em\relax IEEE,
  2003, pp. 7--18.

\bibitem{mittal2014survey}
S.~Mittal, ``{A survey of techniques for improving energy efficiency in
  embedded computing systems},'' \emph{International Journal of Computer Aided
  Engineering and Technology}, vol.~6, no.~4, pp. 440--459, 2014.

\bibitem{del2019hierarchical}
A.~L. D.~M. Martins, A.~H.~L. da~Silva, A.~M. Rahmani, N.~D. Dutt, and F.~G.
  Moraes, ``{Hierarchical adaptive Multi-objective resource management for
  many-core systems},'' \emph{Journal of Systems Architecture}, vol.~97, pp.
  416--427, 2019.

\bibitem{kumar2014chip}
V.~Kumar, ``{On-chip Aging Compensation for Output Driver},'' in
  \emph{International Reliability Physics Symposium (IRPS)}.\hskip 1em plus
  0.5em minus 0.4em\relax IEEE, 2014, pp. CA.3.1--CA.3.5.

\bibitem{chen2003comparison}
T.~Chen and S.~Naffziger, ``{Comparison of adaptive body bias (ABB) and
  Adaptive Supply Voltage (ASV) for Improving Delay and Leakage Under the
  Presence of Process Variation},'' \emph{IEEE Transactions on Very Large Scale
  Integration (VLSI) Systems}, vol.~11, no.~5, pp. 888--899, 2003.

\bibitem{wang2014silicon}
X.~Wang, J.~Keane, T.~T.-H. Kim, P.~Jain, Q.~Tang, and C.~H. Kim, ``Silicon
  odometers: Compact in situ aging sensors for robust system design,''
  \emph{IEEE micro}, vol.~34, no.~6, pp. 74--85, 2014.

\bibitem{igarashi2015digital}
M.~Igarashi, K.~Takeuchi, T.~Okagaki, K.~Shibutani, H.~Matsushita, and K.~Nii,
  ``{An on-die digital aging monitor against HCI and xBTI in 16 nm Fin-FET bulk
  CMOS technology},'' in \emph{European Solid-State Circuits Conference
  (ESSCIRC)}.\hskip 1em plus 0.5em minus 0.4em\relax IEEE, 2015, pp. 112--115.

\bibitem{majerus2017embedded}
S.~Majerus, X.~Tang, J.~Liang, and S.~Mandal, ``{Embedded silicon odometers for
  monitoring the aging of high-temperature integrated circuits},'' in
  \emph{National Aerospace and Electronics Conference (NAECON)}.\hskip 1em plus
  0.5em minus 0.4em\relax IEEE, 2017, pp. 98--103.

\bibitem{sengupta2017estimating}
D.~Sengupta and S.~S. Sapatnekar, ``{Estimating Circuit Aging due to BTI and
  HCI using Ring-Oscillator-Based Sensors},'' \emph{IEEE Transactions on
  Computer-Aided Design of Integrated Circuits and Systems}, vol.~36, no.~10,
  pp. 1688--1701, 2017.

\bibitem{kim2017investigation}
Y.~Kim, H.~Shim, M.~Jin, J.~Bae, C.~Liu, and S.~Pae, ``{Investigation of HCI
  effects in FinFET based ring oscillator circuits and IP blocks},'' in
  \emph{International Reliability Physics Symposium (IRPS)}.\hskip 1em plus
  0.5em minus 0.4em\relax IEEE, 2017, pp. 4C--2.1--4C--2.4.

\bibitem{shakya2015performance}
B.~Shakya, U.~Guin, M.~Tehranipoor, and D.~Forte, ``{Performance optimization
  for on-chip sensors to detect recycled ICs},'' in \emph{International
  Conference on Computer Design (ICCD)}.\hskip 1em plus 0.5em minus 0.4em\relax
  IEEE, 2015, pp. 289--295.

\bibitem{miyake2016temperature}
Y.~Miyake, Y.~Sato, S.~Kajihara, and Y.~Miura, ``{Temperature and Voltage
  Measurement for Field Test Using an Aging-Tolerant Monitor},'' \emph{IEEE
  Transactions on Very Large Scale Integration (VLSI) Systems}, vol.~24,
  no.~11, pp. 3282--3295, 2016.

\bibitem{ali2016accessing}
G.~Ali, A.~Badawy, and H.~G. Kerkhoff, ``{Accessing on-chip temperature health
  monitors using the IEEE 1687 standard},'' in \emph{International Conference
  on Electronics, Circuits and Systems (ICECS)}.\hskip 1em plus 0.5em minus
  0.4em\relax IEEE, 2016, pp. 776--779.

\bibitem{rathore2019lifeguard}
V.~Rathore, V.~Chaturvedi, A.~K. Singh, T.~Srikanthan, and M.~Shafique,
  ``{LifeGuard: A Reinforcement Learning-Based Task Mapping Strategy for
  Performance-Centric Aging Management},'' in \emph{Design Automation
  Conference (DAC)}.\hskip 1em plus 0.5em minus 0.4em\relax ACM, 2019, p. 179.

\bibitem{savanur2015bist}
P.~R. Savanur, P.~Alladi, and S.~Tragoudas, ``{A BIST approach for counterfeit
  circuit detection based on NBTI degradation},'' in \emph{International
  Symposium on Defect and Fault Tolerance in VLSI and Nanotechnology Systems
  (DFTS)}.\hskip 1em plus 0.5em minus 0.4em\relax IEEE, 2015, pp. 123--126.

\bibitem{sai2017cost}
G.~Sai, B.~Halak, and M.~Zwolinski, ``{A cost-efficient delay-fault monitor},''
  in \emph{International Symposium on Circuits and Systems (ISCAS)}.\hskip 1em
  plus 0.5em minus 0.4em\relax IEEE, 2017, pp. 1--4.

\bibitem{sai2018multi}
------, ``{Multi-Path Aging Sensor for Cost-Efficient Delay Fault
  Prediction},'' \emph{IEEE Transactions on Circuits and Systems II: Express
  Briefs}, vol.~65, no.~4, pp. 491--495, 2018.

\bibitem{copetti2016nbti}
T.~Copetti, G.~C. Medeiros, L.~B. Poehls, and F.~Vargas, ``{NBTI-Aware Design
  of Integrated Circuits: A Hardware-Based Approach for Increasing Circuits’
  Life Time},'' \emph{Journal of Electronic Testing}, vol.~32, no.~3, pp.
  315--328, 2016.

\bibitem{sadeghi2015online}
S.~Sadeghi-Kohan, M.~Kamal, J.~McNeil, P.~Prinetto, and Z.~Navabi, ``{Online
  self adjusting progressive age monitoring of timing variations},'' in
  \emph{International Conference on Design Technology of Integrated Systems in
  Nanoscale Era (DTIS)}.\hskip 1em plus 0.5em minus 0.4em\relax IEEE, 2015, pp.
  1--2.

\bibitem{kamalself}
S.~{Sadeghi-Kohan}, M.~{Kamal}, and Z.~{Navabi}, ``{Self-Adjusting Monitor for
  Measuring Aging Rate and Advancement},'' \emph{IEEE Transactions on Emerging
  Topics in Computing}, vol. (preprint), no.~1, pp. 1--1, 2017.

\bibitem{yi2014scan}
H.~Yi, T.~Yoneda, and M.~Inoue, ``{A Scan-Based On-Line Aging Monitoring
  Scheme},'' \emph{JSTS: Journal of Semiconductor Technology and Science},
  vol.~14, no.~1, pp. 124--130, 2014.

\bibitem{jung2017diagnosing}
J.~Jung, M.~A. Ansari, D.~Kim, H.~Yi, and S.~Park, ``{On Diagnosing the Aging
  Level of Automotive Semiconductor Devices},'' \emph{IEEE Transactions on
  Circuits and Systems II: Express Briefs}, vol.~64, no.~7, pp. 822--826, 2017.

\bibitem{vazquez2014error}
J.~Vazquez-Hernandez, ``{Error prediction and detection methodologies for
  reliable circuit operation under NBTI},'' in \emph{International Test
  Conference (ITC)}.\hskip 1em plus 0.5em minus 0.4em\relax IEEE, 2014, pp.
  1--10.

\bibitem{chandra2014monitoring}
V.~Chandra, ``{Monitoring Reliability in Embedded Processors - A Multi-layer
  View},'' in \emph{Design Automation Conference (DAC)}.\hskip 1em plus 0.5em
  minus 0.4em\relax ACM, 2014, pp. 1--6.

\bibitem{masuda2018mttf}
Y.~Masuda and M.~Hashimoto, ``{MTTF-aware design methodology of error
  prediction based adaptively voltage-scaled circuits},'' in \emph{Asia and
  South Pacific Design Automation Conference (ASP-DAC)}.\hskip 1em plus 0.5em
  minus 0.4em\relax IEEE, 2018, pp. 159--165.

\bibitem{vijayan2017workload}
A.~Vijayan, S.~Kiamehr, F.~Oboril, K.~Chakrabarty, and M.~B. Tahoori,
  ``{Workload-aware Static Aging Monitoring and Mitigation of Timing-critical
  Flip-flops},'' \emph{IEEE Transactions on Computer-Aided Design of Integrated
  Circuits and Systems}, vol.~37, no.~10, pp. 2098--2110, 2017.

\bibitem{saliva2015digital}
M.~{Saliva}, F.~{Cacho}, V.~{Huard}, X.~{Federspiel}, D.~{Angot},
  A.~{Benhassain}, A.~{Bravaix}, and L.~{Anghel}, ``{Digital circuits
  reliability with in-situ monitors in 28nm fully depleted SOI},'' in
  \emph{Design, Automation \& Test in Europe Conference \& Exhibition
  (DATE)}.\hskip 1em plus 0.5em minus 0.4em\relax IEEE, 2015, pp. 441--446.

\bibitem{di2019hidden}
G.~Di~Natale, E.~I. Vatajelu, K.~S. Kannan, and L.~Anghel,
  ``{Hidden-Delay-Fault Sensor for Test, Reliability and Security},'' in
  \emph{Design, Automation \& Test in Europe Conference \& Exhibition
  (DATE)}.\hskip 1em plus 0.5em minus 0.4em\relax IEEE, 2019, pp. 316--319.

\bibitem{wang2019aging}
Y.-T. Wang, K.-C. Wu, C.-H. Chou, and S.-C. Chang, ``{Aging-aware chip health
  prediction adopting an innovative monitoring strategy},'' in \emph{Asia and
  South Pacific Design Automation Conference (ASP-DAC)}.\hskip 1em plus 0.5em
  minus 0.4em\relax ACM, 2019, pp. 179--184.

\bibitem{rohbani2018low}
N.~{Rohbani} and S.~{Miremadi}, ``{A Low-Overhead Integrated Aging and SEU
  Sensor},'' \emph{IEEE Transactions on Device and Materials Reliability},
  vol.~18, no.~2, pp. 205--213, 2018.

\bibitem{pachito2012aging}
J.~Pachito, C.~V. Martins, B.~Jacinto, J.~Semi{\~a}o, J.~C. Vazquez,
  V.~Champac, M.~B. Santos, I.~C. Teixeira, and J.~P. Teixeira, ``{Aging-aware
  power or frequency tuning with predictive fault detection},'' \emph{IEEE
  Design \& Test of Computers}, vol.~29, no.~5, pp. 27--36, 2012.

\bibitem{semiao2014aging}
J.~{Semião}, C.~{Leong}, A.~{Romão}, M.~B. {Santos}, I.~C. {Teixeira}, and
  J.~P. {Teixeira}, ``{Aging-aware Dynamic Voltage or Frequency Scaling},'' in
  \emph{Design of Circuits and Integrated Circuits (DCIS)}.\hskip 1em plus
  0.5em minus 0.4em\relax IEEE, 2014, pp. 1--6.

\bibitem{semiao2014performance}
J.~{Semião}, D.~{Saraiva}, C.~{Leong}, A.~{Romão}, M.~B. {Santos}, I.~C.
  {Teixeira}, and J.~P. {Teixeira}, ``{Performance Sensor for Tolerance and
  Predictive Detection Of Delay-faults},'' in \emph{Defect and Fault Tolerance
  in VLSI and Nanotechnology Systems (DFT)}.\hskip 1em plus 0.5em minus
  0.4em\relax IEEE, 2014, pp. 110--115.

\bibitem{cho2015aging}
M.~Cho, C.~Tokunaga, M.~M. Khellah, J.~W. Tschanz, and V.~De, ``{Aging-aware
  Adaptive Voltage Scaling in 22nm high-K/metal-gate tri-gate CMOS},'' in
  \emph{Custom Integrated Circuits Conference (CICC)}.\hskip 1em plus 0.5em
  minus 0.4em\relax IEEE, 2015, pp. 1--4.

\bibitem{ding2016design}
D.~Ding, Y.~Zhang, P.~Wang, H.~Qian, and G.~Li, ``{Design a Delay Amplified
  Digital Aging Sensor Circuit in 65nm CMOS},'' in \emph{International
  Conference on Solid-State and Integrated Circuit Technology (ICSICT)}.\hskip
  1em plus 0.5em minus 0.4em\relax IEEE, 2016, pp. 1449--1451.

\bibitem{li2014robust}
J.~Li and M.~Seok, ``{Robust and In-situ Self-testing Technique for Monitoring
  Device Aging Effects In Pipeline Circuits},'' in \emph{Design Automation
  Conference (DAC)}.\hskip 1em plus 0.5em minus 0.4em\relax IEEE, 2014, pp.
  1--6.

\bibitem{jang2014chip}
B.~Jang, J.~K. Lee, M.~Choi, and K.~K. Kim, ``{On-chip aging prediction circuit
  in nanometer digital circuits},'' in \emph{International SoC Design
  Conference (ISOCC)}.\hskip 1em plus 0.5em minus 0.4em\relax IEEE, 2014, pp.
  68--69.

\bibitem{wang2015aging}
X.~Wang, L.~Winemberg, D.~Su, D.~Tran, S.~George, N.~Ahmed, S.~Palosh,
  A.~Dobin, and M.~Tehranipoor, ``{Aging adaption in integrated circuits using
  a novel built-in sensor},'' \emph{IEEE Transactions on Computer-Aided Design
  of Integrated Circuits and Systems}, vol.~34, no.~1, pp. 109--121, 2015.

\bibitem{kufluoglu2017thermally}
H.~{Küflüoglu}, M.~{Chen}, S.~{Lu}, A.~{Rabindranath}, R.~{Kakoee}, and
  S.~{Hu}, ``{Thermally-aware sensor allocation for real-time monitoring and
  mitigation of FEOL aging in System-on-Chip (SoC) applications},'' in
  \emph{International Reliability Physics Symposium (IRPS)}.\hskip 1em plus
  0.5em minus 0.4em\relax IEEE, 2017, pp. 4C--6.1--4C--6.5.

\bibitem{koneru2015fine}
A.~Koneru, A.~Vijayan, K.~Chakrabarty, and M.~B. Tahoori, ``{Fine-grained aging
  prediction based on the monitoring of run-time stress using DfT
  infrastructure},'' in \emph{International Conference on Computer-Aided Design
  (ICCAD)}.\hskip 1em plus 0.5em minus 0.4em\relax IEEE, 2015, pp. 51--58.

\bibitem{firouzi2015re}
F.~Firouzi, F.~Ye, A.~Vijayan, A.~Koneru, K.~Chakrabarty, and M.~B. Tahoori,
  ``{Re-using BIST for Circuit Aging Monitoring},'' in \emph{European Test
  Symposium (ETS)}.\hskip 1em plus 0.5em minus 0.4em\relax IEEE, 2015, pp.
  1--2.

\bibitem{khan2009self}
O.~Khan and S.~Kundu, ``{A self-adaptive system architecture to address
  transistor aging},'' in \emph{Design, Automation \& Test in Europe Conference
  \& Exhibition (DATE)}.\hskip 1em plus 0.5em minus 0.4em\relax IEEE, 2009, pp.
  81--86.

\bibitem{sadi2017design}
M.~Sadi, G.~K. Contreras, J.~Chen, L.~Winemberg, and M.~Tehranipoor, ``{Design
  of Reliable SoCs With BIST Hardware and Machine Learning},'' \emph{IEEE
  Transactions on Very Large Scale Integration (VLSI) Systems}, vol.~25,
  no.~11, pp. 3237--3250, 2017.

\bibitem{baranowski2015line}
R.~Baranowski, F.~Firouzi, S.~Kiamehr, C.~Liu, M.~Tahoori, and H.-J.
  Wunderlich, ``{On-line Prediction of NBTI-induced Aging Rates},'' in
  \emph{Design, Automation \& Test in Europe Conference \& Exhibition
  (DATE)}.\hskip 1em plus 0.5em minus 0.4em\relax IEEE, 2015, pp. 589--592.

\bibitem{narang2015nbti}
S.~Narang and A.~P. Srivastava, ``{NBTI detection methodology for building
  tolerance with respect to NBTI effects employing adaptive body bias},'' in
  \emph{International Conference on Circuit, Power and Computing Technologies
  (ICCPCT)}.\hskip 1em plus 0.5em minus 0.4em\relax IEEE, 2015, pp. 1--7.

\bibitem{li2018linear}
X.~Li, J.~Qing, Y.~Sun, Y.~Zeng, Y.~Shi, and Y.~Wang, ``{Linear and resolution
  adjusted on-chip aging detection of NBTI degradation},'' \emph{IEEE
  Transactions on Device and Materials Reliability}, vol.~18, no.~3, pp.
  383--390, 2018.

\bibitem{hofmann2010highly}
K.~{Hofmann}, H.~{Reisinger}, K.~{Ermisch}, C.~{Schlünder}, W.~{Gustin},
  T.~{Pompl}, G.~{Georgakos}, K.~v.~{Arnim}, J.~{Hatsch}, T.~{Kodytek},
  T.~{Baumann}, and C.~{Pacha}, ``{Highly Accurate Product-level Aging
  Monitoring in 40nm CMOS},'' in \emph{Symposium on VLSI Technology
  (VLSIT)}.\hskip 1em plus 0.5em minus 0.4em\relax IEEE, 2010, pp. 27--28.

\bibitem{rathore2018himap}
V.~Rathore, V.~Chaturvedi, A.~K. Singh, T.~Srikanthan, R.~Rohith, S.-K. Lam,
  and M.~Shaflque, ``{HiMap: A hierarchical mapping approach for enhancing
  lifetime reliability of dark silicon manycore systems},'' in \emph{Design,
  Automation \& Test in Europe Conference \& Exhibition (DATE)}.\hskip 1em plus
  0.5em minus 0.4em\relax IEEE, 2018, pp. 991--996.

\bibitem{gnad2015hayat}
D.~Gnad, M.~Shafique, F.~Kriebel, S.~Rehman, D.~Sun, and J.~Henkel, ``{Hayat:
  Harnessing Dark Silicon and Variability for Aging Deceleration And
  Balancing},'' in \emph{Design Automation Conference (DAC)}.\hskip 1em plus
  0.5em minus 0.4em\relax IEEE, 2015, pp. 1--6.

\bibitem{fuketa2012adaptive}
H.~Fuketa, M.~Hashimoto, Y.~Mitsuyama, and T.~Onoye, ``{Adaptive performance
  compensation with in-situ timing error predictive sensors for subthreshold
  circuits},'' \emph{IEEE Transactions on very large scale integration (VLSI)
  systems}, vol.~20, no.~2, pp. 333--343, 2012.

\bibitem{bowman201145}
K.~A. Bowman, J.~W. Tschanz, S.-L.~L. Lu, P.~A. Aseron, M.~M. Khellah,
  A.~Raychowdhury, B.~M. Geuskens, C.~Tokunaga, C.~B. Wilkerson, T.~Karnik
  \emph{et~al.}, ``{A 45 nm Resilient Microprocessor Core for Dynamic Variation
  Tolerance},'' \emph{IEEE Journal of Solid-State Circuits}, vol.~46, no.~1,
  pp. 194--208, 2011.

\end{thebibliography}

\section*{Acknowledgments}

Author Fernando Gehm Moraes is supported by FAPERGS (17/2551-0001196-1) and CNPq (302531/2016-5), Brazilian funding agencies. Leonardo Rezende Juracy was financed in part by the Coordenação de Aperfeiçoamento de Pessoal de Nivel Superior - Brasil (CAPES) - Finance Code 001.

%#===========================================================
%# SECTION: BIOGRAPHIES 
%#===========================================================
\section*{Biography}

\noindent \textbf{Leonardo Rezende Juracy}  received a bachelor degree from the Pontifical Catholic University of Rio Grande do Sul (PUCRS), Brazil, in Computer Engineering in 2015, an M.Sc. degree from the PUCRS, Brazil, in Computer Science in 2018, and is currently an Ph.D. student at PUCRS. His research interests include design for testability, fault-tolerant designs, asynchronous designs, resilient designs, networks-on-chip and multi-processor systems-on-chip.\\
\vspace{-6pt}

\noindent \textbf{Matheus Trevisan Moreira}  received a B.S.E degree in Computer Engineering from Pontifícia Universidade Católica do Rio Grande do Sul (PUCRS) in 2011. He also received a M.Sc. degree in Computer Science from the graduate program in Computer Science (PPGCC) at PUCRS in 2012. He has over 50 published articles, in conferences and journals. Also, his Thesis received an award from the Brazilian Society of Microelectronics (SBMICRO) and CEITEC S.A. as the best Ph.D. Thesis in Design, EDA and Test of Integrated Circuits in 2016. He is currently the Director of Technology at Chronos Tech, in San Diego, CA, USA. He has experience in different fields of microelectronics with emphasis on non-synchronous circuits design.\\
\vspace{-6pt}

\noindent \textbf{Alexandre de Morais Amory} received bachelor and master degrees in computer science from the PUCRS University, in 2001 and 2003, respectively. In 2007 he received the Ph.D. in computer science from UFRGS University, Porto Alegre, Brazil. His thesis received an Honorable Mention in the CAPES Thesis Award, in 2008. His professional experience include an internship at Philips Research Laboratories, The Netherlands, in 2005; as a lead verification engineer at CEITEC design house from 2007 to 2009; and as a postdoctoral fellow at PUCRS, from 2009 to 2012. Alexandre is currently a professor at PUCRS University. His research interest include design, test, fault-tolerance, and verification of digital systems, particularly MPSoCs and NoCs.\\
\vspace{-6pt}

\noindent \textbf{Fernando Gehm Moraes} (M'1997--SM'2002) received the Electrical Engineering and M.Sc. degrees from the Universidade Federal do Rio Grande do Sul (UFRGS), Porto Alegre, Brazil, in 1987 and 1990, respectively. In 1994 he received the Ph.D. degree from the Laboratoire d’Informatique, Robotique et Microélectronique de Montpellier), France. He is currently at PUCRS, where he has been an Associate Professor from 1996 to 2002, and Full Professor since 2002. He has authored and co-authored 37 peer refereed journal articles in the field of VLSI design. His primary research interests include Microelectronics, FPGAs, reconfigurable architectures, NoCs and MPSoCs.

\end{document}